\documentclass[%
 reprint,
 superscriptaddress,
 amsmath,amssymb,
 aps,
 prl,
]{revtex4-2}
\usepackage{graphicx}% Include figure files
\usepackage{dcolumn}% Align table columns on decimal point
\usepackage{bm}% bold math
\usepackage{siunitx}
\usepackage{mathtools}
\usepackage{float}
\usepackage[usenames,dvipsnames]{color}
\usepackage{wrapfig}
\usepackage{mathtools}
\usepackage{float}
\usepackage[usenames,dvipsnames]{color}
\usepackage[colorlinks=true, citecolor=blue, linkcolor=blue, urlcolor=blue]{hyperref}
\usepackage{xr-hyper}
\usepackage[printonlyused]{acronym}
\usepackage{layouts}%to print textwidth
\usepackage{url}
\usepackage{xcolor}

\setcitestyle{super}

\usepackage{lipsum}

\newcommand{\Hamburg}{Max Planck Institute for the Structure and Dynamics of Matter, Luruper Chausse 149, 22761 Hamburg, Germany}

\newcommand{\ETH}{Institute for Theoretical Physics, ETH Zurich, 8093 Zurich, Switzerland}

\begin{document}

%\title{Dynamical paramagnetic Meissner effect in bilayer cuprates above $T_c$}
%Dynamical Meissner effect in bilayer cuprates above $T_c$
%Dynamical paramagnetic Meissner effect: 
%A Floquet instability of bilayer superconductors above Tc

\title{Flux-Floquet instability in fluctuating superconductors}

%Flux-Floquet instability in driven fluctuating superconductors
%Giant dynamical paramagnetism in driven fluctuating superconductors
%Pair phase instability leading to  magnetic field generation in a fluctuating bilayer superconductor under strong drive}}
%{Floquet instability of the local inter-layer pair phase in pseudogap $\rm YBa_2Cu_3O_{6+x}$ under strong drive}
%\title{Instability of the inter-layer pair phase in underdoped $\rm YBa_2Cu_3O_{6+x}$ under strong drive}
%\title{Amplification of the interlayer pair phase
%and giant dynamical paramagnetism in %the driven pseudogap phase of 
%stongly driven underdoped $ \rm YBa_2Cu_3O_{6+x}$ }
%Dynamical Meissner effect in driven pseudogap YBa2Cu3O_{6+x}

\author{Marios H. Michael}\email{marios.michael@mpsd.mpg.de}
\affiliation{\Hamburg}
\author{Duilio De Santis}
\affiliation{Physics and Chemistry Dept., Interdisciplinary Theoretical Physics Group, Palermo University, 90128 Palermo, Italy}
\affiliation{\ETH}
\author{Eugene A. Demler}
\affiliation{\ETH}
\author{Patrick A. Lee}
\affiliation{Department of Physics, MIT, 77 Massachusetts Avenue, 02139 Cambridge, MA, USA }
%`E. Segr\`{e}'

\date{\today}

\begin{abstract}
\noindent
In the past decade, photo-induced superconducting-like behaviors have been reported in a number of materials driven by intense pump fields. Of particular interest is the high-$T_c$ cuprate $\rm Y Ba_2 Cu_2 O_{6+x}$ (YBCO), where such effect has been reported up to the so-called pseudogap temperature $T^* \sim 300-400$ K. In a recent tour-de-force experiment, a transient magnetic field which is proportional to and in the same direction of an applied field has been observed outside the sample, suggestive of flux exclusion due to the Meissner effect. In this paper, we present an alternative interpretation of these experiments based on a mechanism that we term the flux-Floquet instability of the sine-Gordon (SG) model. We take as our premise the model of preformed Cooper pairs in the pseudogap phase. Starting from the local superconducting order parameter in equilibrium, we introduce an extended SG model to describe the dynamics of the relative phase between the layers of a bilayer. We demonstrate that a combination of external magnetic field and strong terahertz drive used in experiments by Fava et al.\cite{Sebastian24} results in a novel Floquet instability in the model. This instability leads to currents at the edges of the bilayer formed by defects or grain boundaries, with the current flowing in the opposite direction of the equilibrium screening current, producing a giant paramagnetic magnetization in the same direction as the applied field. We show how this scenario can fit most of the available data. To the extent that this model can account for the data, we conclude that the experiments\cite{Sebastian24} have the important consequence of revealing the presence of local pairing in the pseudogap phase. While the bulk of this paper addresses the experiment on YBCO, our results reveal a new instability in the SG equation that is of fundamental interest, with potential applications such as providing a mechanism for generating large magnetic fields at ultra-fast time scales in Josephson devices.

\end{abstract}

\maketitle
\section{1. Introduction}

Developments in pump and probe experiments have ushered in a new era of ultra-fast control in condensed matter systems\cite{basov_towards_2017,de_la_torre_colloquium_2021,Bao22}. Experimental evidence indicates that pumping materials with an intense laser pulse in the THz frequency region can unlock collective behaviour on ultra-fast time-scales, such as ferro-electricity\cite{Li19}, magnetism\cite{Disa_23,Amano_22,Sriram_22,Siegrist_19}, band structure topology\cite{McIver20}, charge ordering \cite{Alfred24,Pavel20} and superconductivity (SC)\cite{Mitrano16, kappaSalts21,Daniele11, Sambuddha23, Eckhardt24}. A prominent example in this field is the case of driven high-$T_c$ cuprate $ \rm YBa_2Cu_3O_{6+x}$ (YBCO)\cite{Kaiser14,vonHoegen22,taherian24,Marios20,Marios22}. Under a 20 $\rm THz$ (mid-IR) pulse, at $T > T_c$, in the so-called pseudogap phase, time-resolved linear reflectivity measurements have shown signatures similar to that of the low temperature superconducting state for the duration of a few picoseconds after pumping\cite{Ribak23}. 
The effect is observed beyond room temperature, up to the so-called pseudogap temperature $T^* \sim 300-400$ K, and has rightfully attracted a great deal of attention\cite{Orenstein15,Zhang20,Babadi17,Dai21,Dasari18,Zhiyuan20,Pavel22,Okamoto16,Sentef16}. 
%at temperatures above the superconducting transition temperature

The defining property of an SC is the Meissner effect, i.e., perfect diamagnetic screening of the external magnetic field. A key question is whether the reported transient SC-like state also exhibits Meissner-like behavior. A breakthrough experimental study along this direction was published recently\cite{Sebastian24}, examining the response of driven YBCO above $T_c$ in the presence of an external magnetic field. Ref.~\citenum{Sebastian24} indeed reports the detection of a transient magnetic field outside the sample which scales roughly linearly with the applied field and is in the same direction. The transient field's magnitude is of order 10 $ \mu \rm T$ for an applied field of 10 $ \rm mT$ at 100 K. 

\begin{figure*}[t!]
    \centering
    \includegraphics[trim = 1 1 5 0, clip, scale = 1]{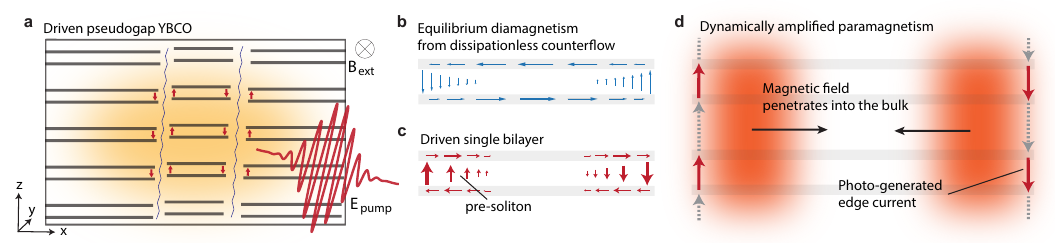}
    \caption{\textbf{Schematic of giant dynamical paramagnetism in driven YBCO}. \textbf{a} Experimental set-up: the bilayer high-$T_c$ cuprate YBCO is pumped with a pulse propagating along the $y$-axis, with the electric field polarised along the $z$-axis, over a spot size of about 100 $\mu$m in diameter, in the presence of an external static magnetic field oriented along the $y$-axis. The sample consists of bilayers lying in the $x$-$y$ plane. It is broken up into finite segments by crystal defects, forming edges. Transient paramagnetic edge currents are generated at the boundaries due to the interplay between diamagnetic screening currents at the edges and Josephson nonlinear dynamics originating from short-range SC coherence in the pseudogap phase. \textbf{b}, \textbf{c} A single blayer at equilibrium, in panel (b), and under strong driving, in panel (c). Panel (b) indicates that above $T_c$, even though long-range coherence and Meissner effect are absent, short-range fluctuating SC coherence within a single bilayer generates a Josephson current (colored in blue) between the two layers, leading to a small static diamagnetism, which we estimate from experiments\cite{Cooper12} to be $\chi_{d} \sim - 10^{-5}$. Panel (c) illustrates that pumping a single bilayer gives rise to soliton-like excitations (dubbed `pre-solitons' here) at the boundaries, characterized by large edge Josephson currents and localized in-plane counterflow currents. The arrows in panels (b) and (c) show the electrical current distribution in the equilibrium and pumped bilayer, respectively. \textbf{d} Upon pumping a YBCO multi-bilayer segment, in the presence of a magnetic field, the large paramagnetic edge Josephson currents in each bilayer (colored red) lead to displacement currents between bilayers (dotted grey). The overall edge current in the multi-bilayer segment generates a magnetic field that penetrates the entire segment at the speed of light, leading to a large magnetic flux inside the segment. The photo-generated edge currents lead to a large paramagnetic susceptibility that can be orders of magnitude larger than the equilibrium diamagnetic response.}
    \label{fig:Sketch}
\end{figure*}

The model presented in Ref.~\citenum{Sebastian24} assumes that the pump pulse induces transient SC generating a diamagnetic current loop, which strongly enhances the magnetic field outside of the illuminated part of the sample. This scenario is analogous to magnetic flux expulsion from the bulk of SCs in the Meissner phase in static systems. By contrast, our analysis relies on a state with short-range superconducting correlations already in equilibrium. Under strong drive, the pump pulse leads to an unstable growth of the relative pair phase between members of a bilayer, which leads to currents at the edges of the bilayer formed by defects or grain boundaries, see Fig.~\ref{fig:Sketch}(a). We find that these currents flow in the opposite direction of the equilibrium screening current, producing a giant paramagnetic magnetization in the same direction as the applied field. A testable consequence of our model is that the paramagnetic signal is induced both inside and outside of the illuminated region.

While the majority of this paper focuses on a specific experimental system, the core of our analysis lies in the identification of a novel instability within the strongly driven sine-Gordon (SG) model. We term this the flux-Floquet instability and demonstrate its potential universality across a broad range of systems with $U(1)$ symmetry; these include excitonic condensates in 2D heterostructures, extended Josephson junctions, and low-dimensional ultracold atomic experiments involving artificial gauge fields~\cite{Likharev86,Ustinov98,Cuevas14,Malomed22,DeSantis23,DeSantis24,DeSantis25,Neuenhahn12,Lovas22,Wybo23}.%By analogy with the equilibrium Meissner effect, Fava et al. have interpreted this phenomenon as originating from flux being expelled from the bulk of the sample.
 To motivate our perspective, we note that in all experiments reported so far, the drive field is applied perpendicular to the plane. The key experimental signatures---out-of-plane optical response and second harmonic generation due to Josephson plasmons~\cite{vonHoegen22,taherian24}---also involve current flow perpendicular to the planes. Since the Meissner effect would require a substantial screening supercurrent to flow both in-plane and out-of-plane, we are motivated to seek an explanation of the recent magnetization data through a model which does not invoke a net in-plane supercurrent flow, but accounting for the dynamical interlayer responses. Our model requires the presence of local pairing, allowing us to define a local pair phase up to the pseudogap temperature. In a separate paper, we demonstrate that the same model can also explain the optical reflectivity and the second harmonic generation data mentioned earlier~\cite{michael2025parametrically}. Therefore, the present paper offers a unified perspective that accounts for all the data related to light-driven YBCO.
These considerations---elaborated further in Sec.~2---lead us to study the behavior of the relative phase $\theta$, which describes the Josephson current between two SC-like layers driven by an intense AC electric field, as modeled by a driven SG equation~\cite{Likharev86,Ustinov98,Cuevas14}, under a magnetic field.
As shown below, we are prompted to consider the one-dimensional SG equation where $\theta$ is driven to oscillate with a large amplitude $\gtrsim \pi$. The SG equation is deep in the nonlinear regime and will be found to exhibit an instability in the response to the external magnetic field.
%As shown below, we are prompted to consider the %highly nonlinear regimeof the
%one-dimensional SG equation %by the following estimate. Using the Josephson relation $\partial_t\theta=2eV/\hbar$ and $V=E \, d_1$,where $d_1$ is the separation between the layers, we find that 
%where $\theta$ is driven to oscillate with a large amplitude $\gtrsim \pi$.  % for the pump field amplitude $E_0 = 2.5$ MV/cm and frequency $\omega_p = 2\pi \times \SI{20}{THz}$ employed experimentally\cite{Sebastian24}. %In other words, the pendulum, which represents the interlayer relative phase, can go near or even over the inversion point  due to the drive.
%The SG equation is deep in the nonlinear regime and will be found to exhibit an instability in the response to the external magnetic field.

In equilibrium, the SG model exhibits the well-known \textit{Pokrovsky-Talapov} (PT) phase transition\cite{Kasper20,Lazarides09}: above a critical magnetic field, topological excitations---solitons---proliferate throughout the system. Out of equilibrium, we find that intense driving triggers an instability, where the time-averaged dynamics of the relative phase exhibits exponential growth. This exponential growth results in the injection of topological defects from the edges of the sample, effectively reaching a PT-like regime at magnetic fields orders of magnitude below the critical field. One of the main achievements of this paper is the discovery of this new instability mechanism, which we term flux-Floquet instability, extensively addressed in Section~3. Beyond its relevance for understanding experiments in pumped YBCO, we emphasize that this dynamical solitonic transition is an important phenomenon in its own right, see Sec.~6 for a detailed discussion on its potential universality across a broad range of systems with $U(1)$ symmetry.

To capture the full dynamics of the electromagnetic field in driven-YBCO experiments, the SG equation is then coupled to Maxwell’s equations. Our analysis shows that, when coupled to light, the onset of the flux-Floquet instability can account for the experimental data both qualitatively and quantitatively. We predict that this instability is accompanied by a giant paramagnetic response---two orders of magnitude larger than and with opposite sign to the diamagnetic response in equilibrium. Beyond identifying a dynamical solitonic transition, the impact of our work lies in providing a microscopic theory that reproduces the experimental data in Ref.~\citenum{Sebastian24}, with strong implications for the nature of the pseudogap phase in high-$T_c$ cuprate SCs. These implications are tied to the assumptions of our theory regarding the existence of short-range SC correlations in equilibrium far above $T_c$.

Below we consider the pumping geometry implemented in the experiments of driven $\rm YBCO$, illustrated in Fig.~\ref{fig:Sketch}(a). The YBCO structure consists of strongly coupled Cu-O layers, separated by distance $d_1 = \SI{4}{	\AA} $, which we refer to as bilayers. These bilayers are separated from each other by a larger distance $d_2 = \SI{8}{\AA}$. Using the Josephson relation $\partial_t\theta=2eV/\hbar$ and $V(t)=E_0 \, d_1 \sin{(\omega_p t)}$, we find that $\theta$ is oscillating at the drive frequency $\omega_p$ with a large amplitude $\gtrsim \pi$ for the pump field strength $E_0 = 2.5$ MV/cm and frequency $\omega_p = 2\pi \times \SI{20}{THz}$ employed experimentally\cite{Sebastian24}. This estimate places the system in the highly nonlinear regime, as mentioned earlier.  In Sec. 2, starting with a minimal set of assumptions concerning the existence of SC phase coherence at equilibrium above $T_c$, we map the electrodynamics of the pseudogap phase of YBCO to the SG model. In Sec. 3, we demonstrate that short-range superconducting coherence within a single bilayer of $\rm YBCO$ leads to a small diamagnetic susceptibility $| \chi_{d} | \sim 10^{-5}$, depicted in Fig.~\ref{fig:Sketch}(b). Upon driving, the flux-Floquet instability is triggered locally in each bilayer as shown in Fig.~\ref{fig:Sketch}(c). Section 4 illustrates once coupled to Maxwell equations different bilayers are capacitively coupled to each other, resulting in dynamically amplified edge currents (and displacement currents), outlined in Fig.~\ref{fig:Sketch}(d). The magnetic field generated by these currents is paramagnetic and it penetrates the entire sample at the speed of light. The photo-generated paramagnetic response is proportional to the equilibrium diamagnetic response but can be orders of magnitude larger in amplitude. In Sec. 5, we discuss in detail how our theory compares with the experimental data. In Sec.~6, we address the instability mechanism's potential universality across a broad range of systems with $U(1)$ symmetry. Finally, we conclude the paper by positioning it in the general landscape of ultrafast probes of electron states and photoinduced phases and its implications on the study of high-$T_c$ SCs. An overview of our theoretical framework is given in the Methods section.

\section{2. Model for driven pseudogap YBCO}

%For YBCO, the key experimental evidences for superconducting-like behavior up to now are the out-of plane optical response and second harmonic generation due to Josephson plasmons\cite{vonHoegen22,taherian24}. Since both phenomena involve current flow perpendicular to the planes, we are motivated to focus our attention to interlayer Josephson currents. 
Our starting assumption is that the phase of a SC order parameter can be defined locally in space and time up to the pseudogap temperature $T^*$, which we identify with a mean-field temperature $T_{MF}$. There is general agreement that phase fluctuations are responsible for the destruction of SC in cuprates above $T_c$\cite{Uemura89,emery1995importance}, driven by the Berezinskii-Kosterlitz-Thouless (BKT) transition\cite{Berezinskii1972DestructionLongRangeII,Kosterlitz1973OrderingMetastability,Jose2013FortyYearsBKT} through the proliferation of vortices in each layer, even though the extent of the fluctuation regime is under debate. While the phase of each individual layer fluctuates wildly, the vortices between members of each bilayer are locked if the Josephson energy of a single bilayer is large enough\cite{fertig2002deconfinement,Egor08}. As a result, the \textit{relative} phase $\theta$ between the layers does not see the vortices and can retain coherence, resulting in equal and opposite (counterflow) supercurrents in the two members of the bilayer above $T_c$. This was demonstrated explicitly in a recent numerical study\cite{Homann24}, which found coherence in the relative phase of bilayers and counterflow supercurrent up to about twice $T_c$. We begin by studying in detail a model where $\theta$ is relatively small at equilibrium and its fluctuations are ignored. We uncover novel phenomena under strong driving, which we demonstrate to occur on a relatively short length-scale near the sample edge, so it is likely that local patches with small $\theta$ mod $2\pi$ are sufficient to see the effect. In 1D, we provide supporting evidence for this intuition in the Methods section by repeating our calculations in the presence of a soliton or an anti-soliton in the initial conditions. We will also offer an argument for the robustness of our mechanism  against thermal fluctuations in 2D. 

In YBCO, the Josephson tunnel coupling between members of the bilayers is much stronger than that between bilayers. We therefore introduce an intra-bilayer Josephson energy $J_c$, while the tunneling between bilayers is set to zero.

\section{3. Dynamical phase diagram of the driven SG model}

\begin{figure*}[t!]
    \centering
    \includegraphics[scale = 1]{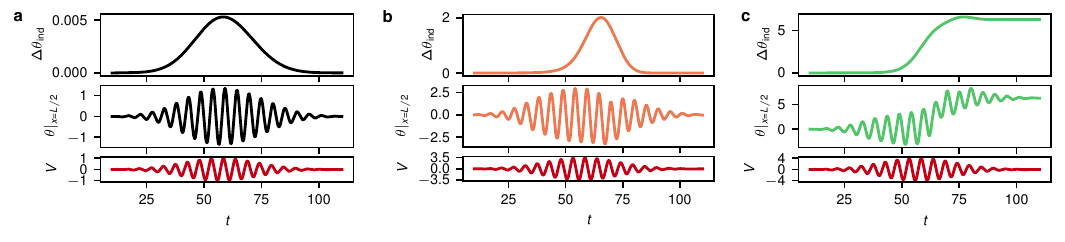}
    \caption{\textbf{Three modes of the time evolution of the phase $\theta$ under drive.} \textbf{a}, \textbf{b}, \textbf{c} correspond to the driving amplitudes $V_0 = 1, 3.5, 4$, respectively, and a modest $q$ value of $0.01$, i.e., a rather weak applied magnetic field. The time profiles of the drive fields are shown in the bottom panels. The middle panels show the time dependence of the phase at the middle of the segment, $x=L/2$. The signal is dominated by an oscillatory component, at the drive frequency, much larger than the maximum equilibrium phase value (which is of order $q$). The top panel shows the phase drop induced by the drive across the entire segment. In \textbf{a}, the induced phase drop is smaller than the equilibrium phase $\approx q$. In \textbf{b}, the phase drop is two orders of magnitude larger than $q$. This amplification is evidence for an instability due to the drive. Note that the oscillator component shown in the middle panel is approaching $\pi$. In \textbf{c}, the phase drop across the sample is $2\pi$, indicating that a soliton has been generated by the drive. The middle panel shows the oscillating phase, whose average has shifted by $2\pi$. The following parameters are used in all plots: $L=100$, $\alpha=0.5$, and $\beta = 10^{-5}$.}
    \label{fig:SGruns}
\end{figure*}
In a single bilayer, the relative phase between the two layers, $\theta(x,y,t)$, obeys the equation\cite{Likharev86,Ustinov98,Cuevas14} (see the derivation in the Appendix 1):
\begin{equation}
\partial_t^2 \theta + \frac{\sigma_{n}}{\epsilon_0} \partial_t \theta  - \frac{c^2 \beta}{1 + \beta} \nabla^2 \theta + \omega_J^2 \sin \theta = \frac{e^* d_1}{\hbar} \partial_t E_p(t).
\label{eq:drivenEOM}
\end{equation}
In Eq.~(\ref{eq:drivenEOM}) we include the electric field due to the pump pulse, which we parametrize as $E_p(t) = E_0 \, e^{- \frac{(t - t_p)^2}{2\sigma^2}} \sin[\omega_p (t - t_0)]$, where $E_0$ ($\omega_p$) is the pump amplitude (frequency) and $t_p$, $\sigma$, and $t_0$ define, respectively, the center time, width, and starting time of the Gaussian envelope. We also introduce a  damping term proportional to the DC normal state conductivity  $\sigma_n$. The intra-bilayer Josephson plasma resonance frequency, $\omega_J$, is given by the Josephson coupling, $J_c$, through the equation $\omega_J^2 = e^* d_1 J_c / ( \hbar \epsilon_0 )$, and is set by the upper plasmon frequency in YBCO: $\omega_J = 2 \pi \times 14$ THz. The Josephson coupling determines the current along the $z$-direction, $j_z = J_c \sin \theta$. We introduce a dimensionless parameter 
\begin{equation}
\beta = \frac{\left(e^*\right)^2 n_s d_1 }{2 \epsilon_0 m c^2},
\label{eq:beta}
\end{equation}
which characterizes the  counterflow supercurrent. It is proportional to the 2D superfluid density of Cooper pairs in each layer, $n_s$; $\epsilon_0$ is the free space permittivity, $m$ is the mass of a Cooper pair, $c$ the speed of light in the material, and $e^* = 2 e$ is the charge of a Cooper pair. We will soon see that its physical manifestation lie in the relation $\beta= - 3 \chi_d $, where $\chi_d$ is the diamagnetic susceptibility due to the SC fluctuations.

%The SG equation, as well as its perturbed counterparts, have been under intense investigation, and a great deal is known in the literature\cite{Cuevas14,Malomed22}. What is less studied is the magnetic response, under strong AC pulses, in the presence of a magnetic field applied in the plane along the $y$ direction, see Fig.\ref{fig:Sketch}(a).
We consider the effect of an applied static field $ B_{\rm{ext}} $ in the plane along the $y$ direction, see Fig.~\ref{fig:Sketch}(a). It is described by the boundary conditions (BC):
\begin{equation}
    \left.\partial_x \theta \right|_{x = 0, L} = \frac{e^* d_1}{\hbar} B_{\rm{ext}},
\label{eq:magnStatic}
\end{equation}
where $x = 0$ ($x = L$) is intended as the left (right) edge of the bilayer segment. Since Eq.~\eqref{eq:magnStatic} depends only on $x$ and the drive is uniform in space, we can consider solutions where $\theta$ is independent of $y$. In Eq.~\eqref{eq:drivenEOM}, the $\nabla$ can be replaced by $\partial_x$ and Eq.~\eqref{eq:drivenEOM} simplifies to a one-dimensional equation: the perturbed SG equation. The SG equation, as well as its perturbed counterparts, have been under intense investigation, and a great deal is known in the literature\cite{Cuevas14,Malomed22}. What is less studied is the magnetic response under strong AC pulses, in the presence of an external magnetic field. We shall examine the interplay between strong pumping and the BC imposed by the external field given by Eq.~\eqref{eq:magnStatic}.

%We are motivated by the following simple estimate. Using the Josephson relation $\partial_t\theta=2eV$ and $eV=E \, d_1$, we expect $\theta$ to increase during the drive and reach values $\gtrsim \pi$, corresponding to the pendulum going over the top and becoming inverted, for the pump field amplitude $E_0 = 2.5$ MV/cm and frequency $\omega_p = 2\pi \times \SI{20}{THz}$ employed experimentally\cite{Sebastian24}. Therefore, under the pumping parameters of the experiment, the SG equation is deep in the nonlinear regime and instabilities are expected. 

\paragraph{Equilibrium pseudogap diamagnetism.} In equilibrium, Eqs.~\eqref{eq:drivenEOM} and \eqref{eq:magnStatic} lead to a diamagnetic response due to the presence of counterflow supercurrents partially screening out the $B_{\rm{ext}}$ field within bilayers, see Fig.~\ref{fig:Sketch}(b). The in-plane counterflow supercurrent is given by the expression: $j_x = \frac{c^2 \epsilon_0 \hbar \beta}{e^* d_1 (1 + \beta)} \left( \partial_x \theta  - \frac{e^* d_1}{\hbar} B_{\rm{ext}} \right)$. In equilibrium, the relation $\partial_x \theta = 0 $ holds in the bulk of the material, leading to a constant diamagnetic current $j_x = - \frac{c^2 \epsilon_0 \beta}{(1 + \beta)} B_{\rm{ext}}$. The resulting diamagnetic susceptibility, due to superconducting fluctuations, is given by $\chi_d = - \frac{\beta}{3}$, estimated from experimental data to be $\beta \sim 10^{-5}$, as described in the Methods (see also Appendix~2).

Near the edges of the sample, the in-plane component of the current goes to zero, being replaced by the Josephson current along the $z$-axis, as shown schematically in Fig.~\ref{fig:Sketch}(b). The penetration depth of the diamagnetic screening current is given by $\lambda=\sqrt{\frac{\beta}{1+ \beta}}\frac{c}{\omega_J}$. Rescaling the $x$ coordinate by $\lambda$ and the $t$ coordinate by $\omega_J^{-1}$, we obtain the perturbed SG equation (and its BC) in dimensionless form:
\begin{align}
    \partial_t^2 \theta + \alpha \partial_t \theta  - \partial_x^2 \theta + \sin \theta = & \; \partial_t V, \label{eq:rescaledEOM} \\
    \left. \partial_x \theta\right|_{x = 0, L} = & \; q ,
\label{eq:rescaledBC}
\end{align}
where we define the driving term $V(t) = V_0 \, e^{- \frac{(t - t_p)^2}{2\sigma^2}} \sin[\omega_p (t - t_0)]$ with amplitude $V_0=\frac{e^* d_1 E_0}{\hbar \omega_J}$, the damping coefficient $\alpha = \frac{\sigma_{n}}{\epsilon_0 \omega_J}$ and the boundary field $q = \frac{e^* d_1 \lambda B_{\rm{ext}}}{\hbar}$. Motivated by the experimental set-up\cite{Sebastian24}, we choose $\omega_p = 1.2 $, $t_p=t_0+3\sigma$, $\sigma = 6 \pi/\omega_p $ and $t_0=10$. 

\begin{figure*}[t!]
    \centering
    \includegraphics[scale = 1]{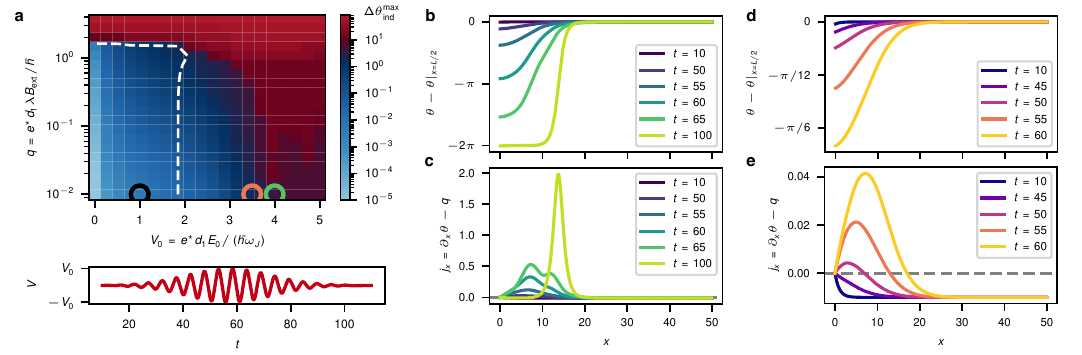}
    \caption{\textbf{Dynamical phase diagram of the driven SG model.} \textbf{a} Heatmap of the maximum photo-induced phase difference across the entire segment, $\Delta \theta^{\rm{max}}_{\rm{ind}} $, in the ($V_0$, $q$) parameter space (upper plot) and time trace of the pump pulse with amplitude $V_0$ (lower plot). The PT transition is observed in the unpumped ($V_0 = 0$) limit for high enough magnetic field. A rather sharp transition occurs between a soliton regime, marked by red, and a pre-soliton regime, marked by blue. This transition separates the points indicated by the orange and green circles, the behavior of which was illustrated in Fig.~\ref{fig:SGruns}. Note that the boundary is almost vertical and extends to very small $q$. This indicates a large amplification of the phase due to the drive. The dashed white line indicates where the maximum phase drop has reached the value $2q$---a rough indicator for the onset of unstable growth of the phase $\theta$ (see the text for a more detailed explanation). \textbf{b} Snapshots of the phase and \textbf{c} snapshots of in-plane current for the parameters $V_0 = 4$ and $q = 0.01$, representative of the soliton regime, corresponding to the point annotated with a green circle in panel (a). The small localized phase profile near the edge, present in equilibrium, monotonically grows upon driving until a full $2 \pi$ soliton enters the system. We find a photo-induced sign change of the in-plane current, as shown in panel (c). \textbf{d} Snapshots of the phase and \textbf{e} snapshots of in-plane current for the parameters $V_0 = 3.5$ and $q = 0.01$, taken in the pre-soliton regime, corresponding to the point annotated with an orange circle in panel (a). A boundary-localized feature, resembling a soliton but with much smaller amplitude, develops during pumping. This induces the sign change of the in-plane current shown in panel (e). The following parameters are used in all plots: $L=100$, $\alpha=0.5$, and $\beta = 10^{-5}$. In both panels (b) and (d), the difference $\theta - \theta_{x=L/2}$ is shown in order to eliminate the spatially uniform, fast-oscillating component of the phase. The dashed gray line in panel (e) indicates $j_x = 0$.}
    \label{fig:SGoverview}
\end{figure*}
\paragraph{Flux-Floquet instability and Josephson solitons.} To describe the effect of an external drive in the presence of an applied magnetic field, we solve Eqs.~\eqref{eq:rescaledEOM} and \eqref{eq:rescaledBC} numerically. In Fig.~\ref{fig:SGruns}, we show the results that exhibit three distinct behaviors for increasing driving amplitudes $V_0$, with a small value of $q=0.01$. In each case, we display the time-dependence of the drive $V(t)$ in the bottom panel. The middle panel illustrates the phase $\theta (t)$ at $x=L/2$---that is, in the middle of the 1D segment. The top panel shows the quantity $\Delta \theta_{\rm{ind}} = \left.\theta\right|_{x=0}^{x=L} - \left. \theta_{\rm{eq}} \right|_{x=0}^{x=L}$, i.e., the drop of the phase across the segment (after subtracting the equilibrium phase drop) as a function of time. Fig.~\ref{fig:SGruns}(a) is obtained with a modest driving amplitude $V_0=1$. In the middle panel, we observe that $\theta$ exhibits sinusoidal oscillation at the drive frequency, as expected for a modest drive far from resonance. Since the boundary condition has a very weak effect for small $q$, this response is essentially uniform along $x$. The top panel displays that the overall drop of the phase across the sample reaches a value smaller than $q$, which is again as expected. Fig.~\ref{fig:SGruns}(b) demonstrates a dramatic change of the response as the driving amplitude is increased to $V_0=3.5$. In the middle panel, we see that the phase is still oscillatory at the drive frequency, but the amplitude has almost reached $\pi$---the pendulum is almost inverted. The oscillatory part remains weakly dependent on $x$ (not shown here). Remarkably, the phase drop across the sample (see the top panel) reaches a value near $2$, which is two orders of magnitude larger than $q$. Examining the spatial dependence of the solution in Fig.~\ref{fig:SGoverview}(b) and (c), we find that this behavior arises because the phases near the edges are strongly amplified, causing a large phase drop across the sample. The localization of the phase increase near the edge explains why it is not visible in the middle of the segment, as shown in the middle panel.

For further increase of the amplitude to $V_0=4$, as shown in the middle panel of Fig.~\ref{fig:SGruns}(c), the amplitude of the oscillatory part of the phase goes beyond $\pi$, but in contrast to the previous cases, its average value shows a jump of $2\pi$. This is reflected in a jump of the overall phase by $2\pi$ in the top panel. The solution near the edge---see Fig.~\ref{fig:SGoverview}(d) and (e)---shows that, as the phase gets amplified, a soliton is launched and it propagates into the bulk of the sample. Depending on the duration of the driving and the damping rate, one or more solitons can be created. We refer to this as the \textit{soliton regime}. Details of the spatial and temporal development of the soliton and its direction of propagation can be found in Appendix 4.

We performed extensive runs in the parameter space $(q, V_0)$. The results are shown in Fig.~\ref{fig:SGoverview}(a). 
The three cases described earlier are shown by the black, orange and green circles, corresponding to the color code in Fig.~\ref{fig:SGruns}. In Fig.~\ref{fig:SGoverview}(a) the red region corresponds to the soliton regime. The dark blue region is where the phase shows strong enhancement, associated with an edge-instability due to the drive. We dub this the \textit{pre-soliton regime}. For further reduction of the drive amplitude $V_0$, we reach the light blue region, where the phase is stable and remains small and less than $q$. As a point of reference, we mention that in the absence of drive ($V_0=0$), there is a phase transition from a commensurate phase for small $q$---where the phase is pinned to zero in the bulk---to an incommensurate phase for large $q$---where the phase develops a set of solitons. This is the well-known the PT transition\cite{Kasper20,Lazarides09} and it happens at the value $q=2$ (see the horizontal part of the dashed white line for large $q$ and small $V_0$).

It is important to note that the boundary between the soliton and pre-soliton regimes in Fig.~\ref{fig:SGoverview} is nearly vertical and extends down to very small $q$. This indicates that phase amplification, and the resulting soliton generation, originates from a dynamically driven instability: the Flux-Floquet instability. One consequence is that the soliton injection, which normally occurs at $q=2$, is now driven down to dramatically small $q$ (i.e., small magnetic fields), with onset at a critical value of $V_0$. To the best of our knowledge, this instability of a small boundary condition under strong driving of the SG model has not been previously described in the extensive literature devoted to this equation. While somewhat related solitonic instabilities have been discussed previously, for example in boundary-driven settings involving nonlinear supratransmission~\cite{geniet2002energy,dienst2013optical} and in post-quench dynamics~\cite{GREENE1999423,Neuenhahn12}, the mechanism identified here is distinct in that it arises from the interplay of a weak boundary twist and a spatially homogeneous strong drive. This constitutes the main mathematical result of this paper.

\begin{figure*}[t!]
    \centering
    \includegraphics[scale = 1]{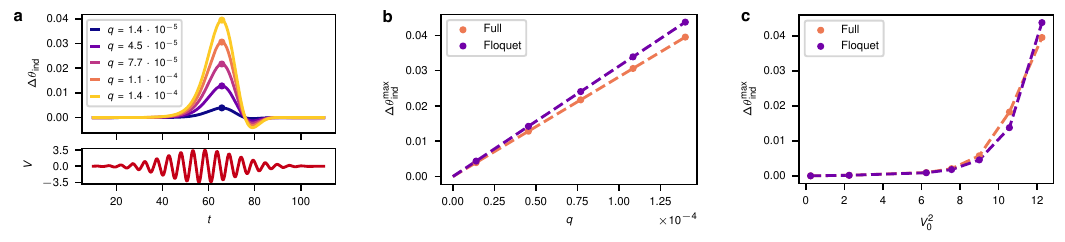}
    \caption{\textbf{Pre-soliton regime of the driven SG model.} \textbf{a} Time traces of the photo-induced phase difference across the entire segment, for the full nonlinear model, at various external magnetic fields (upper plot) and the corresponding pump pulse (lower plot). Circles are used in the upper panel to the denote the maxima of the different curves, all obtained for $V_0 = 3.5$. \textbf{b} Maximum photo-induced phase difference vs. the magnetic field, at $V_0 = 3.5$. The orange (purple) points are obtained by numerically solving the full nonlinear (Floquet effective) model. Despite the strongly nonlinear dynamics, the dependence on the external magnetic field is linear, and the two models show remarkable quantitative and qualitative agreement. \textbf{c} Maximum photo-induced phase difference as a function of fluence, $V_0^2$, at $q = 1.4 \times 10^{-4}$. The orange points result from numerical integration of the full nonlinear model, while purple points correspond to the effective Floquet model. An initial linear increase for small fluences is then followed by an exponential increase beyond a critical amplitude. The two models are in clear quantitative and qualitative agreement here as well. The following parameters are used in all plots: $L=100$, $\alpha=0.5$, and $\beta = 10^{-5}$.}
    \label{fig:SGresults}
\end{figure*}
As we see from Fig.~\ref{fig:SGruns}(a) and (b), the onset of instability can be characterized by a phase drop across the sample which exceeds $q$. In Fig.~\ref{fig:SGoverview}(a), the dashed white line indicates where the maximum phase drop has reached the value $2q$---a rough indicator for the onset of unstable
growth of the phase $\theta$. Below we develop an analytic theory based on decomposing $\theta(x,t)$ into a homogeneous oscillatory component and a slowly varying component $\theta'(x,t)$ satisfying the BC. For small $q$, we show that an instability occurs in the slow component when the amplitude $A$ of the oscillatory component satisfies the condition $A \gtrsim A_0$, where $A_0$ is the first zero-crossing of the Bessel function $J_0(A)$, i.e., $A_0 \approx 2.4 \approx 2\pi/3$. This prediction yields a vertical line in Fig.~\ref{fig:SGoverview}(a) which is consistent with the vertical part of the dashed white line in Fig.~\ref{fig:SGoverview}(a) for small $q$.

In Fig.~\ref{fig:SGoverview}(b) and (c), the phase and in-plane current spatial profiles are plotted for different times during the drive, for a ($V_0$, $q$) pair representative of the soliton regime, see the green circle in Fig.~\ref{fig:SGoverview}(a). We observe that the small localized phase profile near the edge, present in equilibrium, monotonically grows upon driving until it becomes a full $2 \pi$ soliton, which then enters the system. The typical phase and current profiles observed in the dark blue region of the phase diagram, i.e., where full solitons are not excited, are depicted in Fig.~\ref{fig:SGoverview}(d) and (e). We dubbed this the pre-soliton regime, as the boundary-localized dynamical feature resembles a soliton but has a significantly smaller amplitude. We find that experiments\cite{Sebastian24} are deep in the pre-soliton regime, with $V_0 \approx 3.5$ and $q \approx 10^{-4} - 10^{-5}$ such that $\theta(t=0) \ll 1  $ and $\Delta \theta^{\rm{max}}_{\rm{ind}} \lesssim 0.1$.

We emphasize that the instability initially takes place at the edge. The phase's growth produces currents which serve as an antenna localized near the edge that drives an enhanced $B$ field decaying into the bulk. It is also worth noting that in the unstable regime, the effective Josephson energy (denoted $\eta$ in what follows) becomes negative, causing the current along $z$, and also along $x$---shown in Fig.~\ref{fig:SGoverview}(c) and (e)---to reverse direction relative to the equilibrium diamagnetic current. As a result, the generated $B$ field is \textit{paramagnetic} and can be orders of magnitude larger than the equilibrium screening field. This is the key physical consequence of our work and has important implications for interpreting experiments on YBCO.

In Fig.~\ref{fig:SGresults}(a), we plot the $\Delta \theta_{\rm{ind}} = \left.\theta\right|_{x=0}^{x=L} - \left. \theta_{\rm{eq}} \right|_{x=0}^{x=L}$ profiles as function of time, for several values of $q$ in the experimental range, at $V_0 = 3.5$. We find that $\Delta \theta_{\rm{ind}}$ is positive and increases during pumping, before returning back to equilibrium after the pump pulse is gone. The full solution of Eqs.~\eqref{eq:rescaledEOM} and \eqref{eq:rescaledBC} shows a linear dependence of $\Delta \theta^{\rm{max}}_{\rm{ind}}$ vs. $q$ in the pre-soliton regime, for fixed $V_0$, see the orange points in Fig.~\ref{fig:SGresults}(b). The orange curve in Fig.~\ref{fig:SGresults}(c) illustrates the behavior of $\Delta \theta^{\rm{max}}_{\rm{ind}}$ with respect to pump fluence, $V^2_0$, for fixed $q$. Here we observe a linear increase, similar to an inverse Faraday effect\cite{Hamed21}, for small fluences $V^2_0 < 10$. Beyond a critical value of $V^2_0 \approx 10 $, we find an exponential growth of the maximal photo-induced phase difference.

\paragraph{Floquet effective model for the pre-soliton regime.} To better understand the physics of the pre-soliton regime, we decompose the solution of our nonlinear problem into two parts: $\theta(x,t)=\tilde{\theta}(t)+\theta'(x,t)$, where $\tilde{\theta}(t)$ obeys the driven damped pendulum equation of motion
%with  BC $\partial \tilde{\theta}(x=0,L)/\partial x =0$
\begin{equation}
    \partial_t^2\tilde{\theta} + \alpha \partial_t \tilde{\theta} + \sin \tilde{\theta} = \partial_t V(t) ,
    \label{eq:pendulum}
\end{equation}
%which satisfies the homogeneous BC and is therefore constant is space,
with BC $ \partial_x \tilde{\theta}|_{x = 0, L} = 0$, and $\theta'(x,t)$ satisfies the BC given by Eq.~\eqref{eq:rescaledBC}. % inhomogeneous BCs due to the external magnetic field.
In the small $q$ limit relevant for the pre-soliton regime, the condition $\theta' \ll 1$ holds even if the pump is strongly nonlinear and $\tilde{\theta}$ is not small. In this case, we can expand the nonlinear term in Eq.~\eqref{eq:rescaledEOM} as $\sin \theta \approx  \theta' \cos{\tilde{\theta}} $. We begin by factorizing this product and replace  $\cos{\tilde{\theta}} $ by $\langle \cos{\tilde{\theta}} \rangle$, its average over several cycles at the pump frequency. We obtain the following $\it{linear}$ equation, where the drive is encoded in a modified restoring force $\eta$, subject to the same BC
\begin{align}
    \partial_t^2 \theta' + \alpha \partial_t \theta' - \partial_x^2 \theta' + \eta \theta' =& \; 0 , \label{eq:effectiveEOM}\\
    \left. \partial_x  \theta'\right|_{x = 0, L} =& \; q . \label{eq:effectiveBC}
\end{align}
%where $\eta(A(t))$, the effective Josephson coupling, is computed through  Floquet perturbation theory described in the Methods section and given by:
By  approximating $\tilde{\theta} \approx A \sin{(\omega_p t)}$, where $A(t)$ is a slow-varying envelope, we find that $\eta(A) \approx \langle \cos{ \left[ A \sin{(\omega_p t)} \right] } \rangle \approx J_0(A)$ where $J_0$ is the Bessel function.  %n$-th Bessel function of the first kind. 
Importantly, this average can be negative since $J_0$ can reach a minimum value of $\sim - 0.4$. Thus the oscillator in Eq.~\eqref{eq:effectiveEOM} can be in the unstable regime, resulting in the  amplification found in the exact solution. It is clear that the average $\langle \cos \left[ A \sin{(\omega_p t)} \right] \rangle $ does not depend on the phase of the drive, which we have set to be zero. Therefore, the sign of the effect does not depend on the phase of the drive, which is in agreement with our exact numerical solution. In the Methods (see also Appendix~3), we derive a more accurate expression for $\eta(A)$ using a Floquet expansion.

%This is the reason for our choice of this value to generate the dashed curve in Fig.~\ref{fig:SGoverview}(a) to estimate the onset of the instability.
Note that in this approximation, instability occurs when $J_0(A)$ changes sign. This occurs for $ A = A_0 \approx 2.4 \approx 2\pi/3$. It is also worth pointing out that, at the critical value $A_0$, the phase has not reached the inversion point. Thus the instability only requires the phase to sample a highly nonlinear regime, rather than reaching the inversion point.% that significantly improves the agreement with the exact numerical solution.%given by
Remarkably, there is very close agreement between the full nonlinear solution (shown by the orange points) and that of the Floquet effective model (shown by the purple points) in both Fig.~\ref{fig:SGresults}(b) and (c). 
The effective model gives us an intuitive way to understand the numerical results: even though the homogeneous oscillations caused by the drive are very large, the effective dynamics for $\theta'$ are linear to a very good approximation, since $q$ is small. Due to this linearity, the photo-induced response is proportional to $q$. On the other hand, for sufficiently large amplitudes $V_0$, %Eq.~\eqref{eq:pendulum} leads to $ A \sim \pi $, causing 
$\eta(A)$ in Eq.~\eqref{eq:effectiveEOM} becomes negative, thereby triggering an instability and exponential growth of $\theta'$.

In the Methods, we repeat the SG calculations in the presence of a soliton or an antisoliton in the initial conditions. Our results indicate that, as long as these excitations are located more than a few penetration depths ($\lambda < \SI{100}{nm} $) away from the bilayer segment's edges, the discovered mechanism is not affected by their presence.
\begin{figure*}[t!]
\centering
\includegraphics[scale = 1]{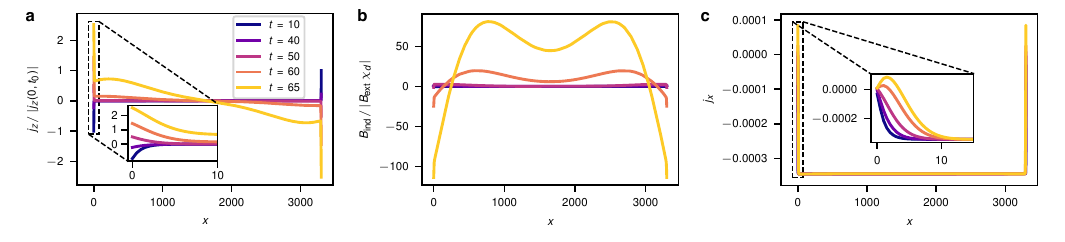}
\caption{\textbf{Photo-induced instability in a multi-bilayer segment.} \textbf{a} Snapshots of the effective current along the $z$-direction, $j_z$, in units of its magnitude at $x=0$ before pumping, $\left|j_z(0, t_0)\right|$. The inset plot zooms on the $0 \leq x \leq 10$ region. When reaching the unstable regime in the presence of strong enough driving, $j_z$ can change sign and become paramagnetic even though $|\theta'|$ monotonically increases. \textbf{b} Snapshots of the average magnetic field induced within a unit cell, $B_{\rm{ind}}$, normalized by the equilibrium diamagnetic field amplitude, $|B_{\rm{ext}} \chi_d|$. The field originates from edge currents, and it propagates from the system's boundaries inwards at the speed of light in the YBCO crystal. \textbf{c} Snapshots of the dimensionless in-plane current $j_x$, showing a dynamical behavior analogous to that highlighted for the pre-soliton currents in a single bilayer shown in Fig. \ref{fig:SGoverview}(e). The inset zooms in to the $0 \leq x \leq 15$ region. All three plots are obtained using the Floquet effective model extended to multilayers (see Methods), for the following realistic (normalized) parameter set: $L=3300$, $V_0 = 3.5$, $q=2.3 \times 10^{-4}$, $\left| \chi_d \right| = 10^{-5}$, $\alpha_1=0.4$, $\alpha_2=6$, which denote damping in the intra- and inter-bilayer regions (see Methods). The legend in panel (a) applies to panels (b) and (c) as well, where time is rescaled to $1/\omega_J$.}
%Oscillations at the edges of the segment, indicating the emission of magnetic field by the strip, can be appreciated.
%[THIS SENTENCE SEEMS REDUNANT> PL]
\label{fig:multi-bilayer}
\end{figure*}
More generally, we argue that our results remain robust in the presence of thermal fluctuations $\theta(x,y,t)$ and is applicable up to the pseudogap temperature $T^*$. Due to the Josephson energy, which acts as an external field in the XY model, there will be patches where the equilibrium $\theta$ is pinned to the vicinity of multiples of $2\pi$. If the correlation length in these patches is longer than the penetration depth $\lambda$, a 2D version of the scaled Eq.~\eqref{eq:rescaledEOM} will hold.  %We can do a similar decomposition into $\tilde{\theta}$ and $\theta'$. %For small $q$ $\theta'$ obeys an equation similar to \eqref{eq:effectiveEOM} 
Assuming the fluctuations evolve slower than the pump duration, we can take a snapshot as  initial conditions for $\theta$ and do a similar decomposition into $\tilde{\theta}$ and $\theta'$. %the rapidly oscillating field $\tilde{\theta}(x,y,t)$. We can do a similar decomposition into $\tilde{\theta}$ and $\theta'$.
For small $q$, $\theta'$ satisfies Eqs.~\eqref{eq:effectiveEOM} and \eqref{eq:effectiveBC}, with  $\eta$ given by $\langle \cos \tilde{\theta}(x,y,t) \rangle$. Upon temporal and space average, the sensitivity to initial condition will average out, and $\eta$ can still become negative, leading to exponential growth of $\theta'$ and a similar phenomenology to the one described in this section. 

Our results show a significant dependence on the dissipation coefficient, $\alpha$. Since the large enhancement of $\Delta \theta^{\rm{max}}_{\rm{ind}}$ is triggered by an instability, $\alpha$ directly impacts the magnitude of the effect by modifying  its growth rate. In Appendix~4, we illustrate this dependence by presenting simulations for different values of $\alpha$. Interestingly, while small $\alpha$ leads to a large enhancement of the effect, in the limit of too small of a dissipation, the emergence of oscillatory modes such as low-amplitude breathers\cite{DeSantis23,DeSantis24,DeSantis25} can complicate the simple picture presented above, leading to a more oscillatory behaviour in the $\Delta \theta_{\rm{ind}}$.

For completeness, in Appendix~4, we present a number of space-time contour plots for the evolution of the phase and its gradient, $\theta$ and $\partial_x \theta$, together with animations of the same numerical runs, to further illustrate the strongly nonlinear dynamics discussed here.

\section{4. Pumped multi-bilayer pseudogap YBCO} 
\label{sec:bilayer}
In this section, we present results for the pumped multi-bilayer case. Although the Josephson coupling between the bilayers is zero, they remain capacitively coupled through Maxwell equations. We simulate the dynamics in a finite segment in the $x$ direction, with radiative BC, see the Methods for a detailed account of the equations of motion. This allows to track both the induced magnetic field inside a finite pumped YBCO strip, as well as the emitted radiation outside of the strip. In the Appendix~5, we further explore the dependence on dissipation.

\paragraph{Photo-excited edge currents in the multi-bilayer system.} At the microscopic level, we attribute the origin of the photo-induced giant paramagnetic response to the flux-Floquet instability of the SG model discussed above. More specifically, within our Floquet effective theory, we established that the magnitude of the phase inside the bilayer, $|\theta'|$, exponentially grows when ${\eta(A) < 0}$. At the same time, the effective current along the $z$-direction is given by $j_z = \eta(A) J_c \sin \theta'$, which changes sign and become paramagnetic when $\eta < 0$, as shown in Fig.~\ref{fig:multi-bilayer}(a), even though $|\theta'|$ monotonically increases during pumping. The latter current creates a displacement current of similar magnitude across different bilayers due their capacitive and resistive coupling. This leads to paramagnetic edge currents, with opposite signs at the two edges. Remarkably, the key difference between the multi-bilayer and the single bilayer is that the space between bilayers now acts as a wave-guide, so that the displacement current and the magnetic field it produces propagate into the bulk at the speed of light in the YBCO crystal from the system's boundaries inwards, see Fig.~\ref{fig:Sketch}(d) and 
Fig.~\ref{fig:multi-bilayer}(b). In Fig.~\ref{fig:multi-bilayer}(c), we show the corresponding in-plane current $j_x$, whose dynamical structure is consistent with that of the pre-soliton currents identified previously for a single bilayer. 

\begin{figure}
    \centering
    \includegraphics[scale = 1]{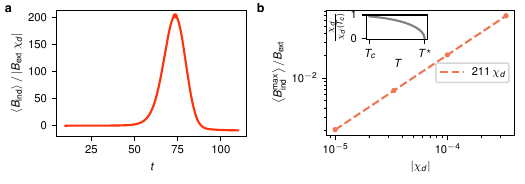}
    \caption{\textbf{Time dependent induced magnetic field under strong pumping} . \textbf{a} Average induced magnetic field, $\left\langle B_{\rm{ind}} \right\rangle$, normalized by the equilibrium diamagnetic field amplitude, $|B_{\rm{ext}} \chi_d|$, as a function of time. Here the symbol $\left\langle ... \right\rangle$ indicates averaging over a segment of length $L$. The obtained peak magnetic response is $\chi_{\rm{ind}} \sim - 100 \, \chi_{d}$, see the circle. The parameter values for this numerical run are given in Fig.~\ref{fig:multi-bilayer}. \textbf{b} Dependence of the (normalized) peak magnetic response, $\left\langle B^{\rm{max}}_{\rm{ind}} \right\rangle / B_{\rm{ext}} $, versus the equilibrium susceptibility $\chi_d$, showing proportionality. Since $\chi_d$ is proportional to the local superfluid density $n_s$, which goes to zero in a BCS way near $T^* \approx T_{MF}$ as sketched in the inset, we expect the signal to decrease to zero near $T^*$, in agreement with experiment. Here we choose $L = 18 \, c / \omega_J $, $B_{\rm{ext}} = \SI{10}{m T}$, $V_0 = 3.5$, $\alpha_1=0.4$, and $\alpha_2=6$.}
    \label{fig:multi-bilayer2}
\end{figure}

Even though the current at the edge is similar in magnitude with the single bilayer case, the upshot is that the magnetic field extends into the bulk, therefore greatly increasing the flux that is generated. This mechanism is crucial for us to obtain a large enough effect to compare with experiment.  Using realistic parameters, the induced field averaged inside and between bilayers, $B_{\rm{ind}}(x)$, becomes $O(100)$-times larger than the equilibrium diamagnetic field amplitude, $|B_{\rm{ext}} \chi_d|$. This is shown in Fig.~\ref{fig:multi-bilayer2}(a). We also show that the magnitude of the induced magnetic field scales linearly with the equilibrium diamagnetic susceptibility $|\chi_d|$ over more than an order of magnitude, see Fig.~\ref{fig:multi-bilayer2}(b). Since $\chi_d$ is proportional to the local superfluid density $n_s$, it is expected to go to zero in a BCS way near $T^* \approx T_{MF}$ as sketched in the inset. The induced magnetic field amplitude is expected to show this temperature dependence, in agreement with experiment. The traveling-wave character of our mechanism also results in a peculiar segment-size dependence that is addressed in the Appendix~5. For completeness, in our simulations, we also track the magnetic fields emitted outside the pumped region in the infinite strip geometry, which we discuss in the Methods. In general, due to flux conservation, any short-lived response in pumped YBCO with a definite sign, either diamagnetic or paramagnetic, will result in a bimodal emission pattern in the far-field, carrying both positive and negative signs, averaging to zero in time. In order to make connection with experiment, it is necessary to modify this geometry to take into account the presence of a mask blocking the pump for $z > 0$. This will be discussed in the next section.

%Because of this observation, we conclude that in this geometry, the radiated magnetic field outside of the strip  between $x=0$ and $L$ where the multilayers are located do not give a net magnetic field. The connection to experiment requires a modification of this geometry, as discussed in the next section. %we attribute the origin of the paramagnetic response measured in Ref.~\citenum{Sebastian24} to the magnetic field generated inside the material, rather than to emitted radiation.

\section{5. Relation to experiments}

\begin{figure*}
    \centering
    \includegraphics[scale = 1]{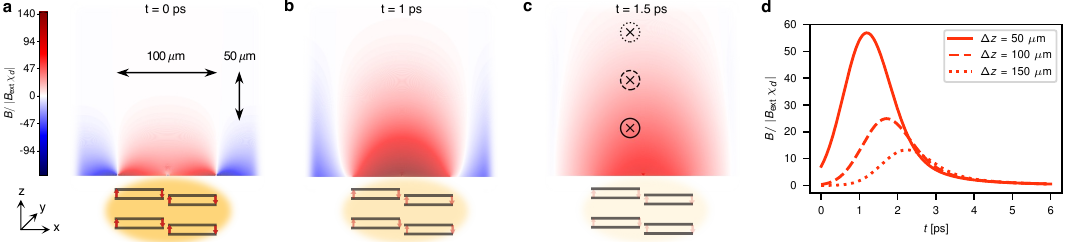}
    \caption{\textbf{Electromagnetic wave propagating from the pumped region.} Geometry of the detection method in experiments. A mask is used to protect the magnetic field detector from the pump pulse that covers the area $z>0$. This area is shown in panels (a), (b), and (c). Illuminated defects and edges will give rise to a paramagnetic signal, propagating at the speed of light in the material and reaching the region underneath the detector. To illustrate this, we report snapshots of the magnetic field distribution beneath the mask at time \textbf{a} $ t = 0$ ps, \textbf{b} $ t = 1$ ps, and \textbf{c} $ t = 1.5$ ps. The axes, physical scales, and colorbar from panel~(a) apply to panels~(b) and (c) as well. This calculation is performed on a two-segment system, pumped over $100 \; \mu$m along the $z$ direction. We also assume a Gaussian-like induced response centered at $t = 0$ and extending in time over $2$ ps. \textbf{d} Magnetic field profile $B$ versus time at three selected locations: $\Delta z = 50 \; \mu$m, solid line; $\Delta z = 100 \; \mu$m, dashed line; $\Delta z = 150 \; \mu$m, dotted line. This calculation is produced for the observation coordinate $x = 40 \; \mu$m, see panel~(c) for a sketch, under the same geometry and induced response of the previous panels. Throughout the figure, the magnetic field $B$ is scaled by the equilibrium diamagnetic field amplitude.}
    \label{fig:multi-bilayer3}
\end{figure*}

In the experiment, a mask is placed over the magnetic field detector to protect it from the high-intensity pump field. The edge of the mask is parallel to the long axes of the bilayer [$x$-axis in Fig.~\ref{fig:multi-bilayer3}(a)] and covers the area $z>0$. This geometry is different from what is considered in the last section, where the YBCO is an infinitely long strip running along $z$. The important difference is that now the radiation generated by the edge current can escape into the $z>0$ region and reach the detector. As mentioned earlier, the sample is broken up into domains separated by cracks and defects. The typical domain size is roughly $\rm 10 \; \mu m$. A transient current is generated at the edge of each domain, as sketched by the red arrows in Fig.~\ref{fig:multi-bilayer3}(a). Each domain constructively contributes to the emitted radiation in the $z > 0$ area and we expect the net result to be similar to that of a single domain with the size of the spot, but we have not pursued detailed modeling of this correspondence.

As a test case, we consider two domains spanning a total length of $100 \; \rm{\mu m}$, corresponding to the spot size. We investigate the emitted magnetic field distribution in the shielded region ($z > 0$) due to time-dependent 2D edge currents with a Gaussian time profile of duration $\sim 2$ ps (centered at $t=0$). For simplicity, the layers are assumed to be infinite along $y$, ensuring translation invariance in that direction. A time-dependent sheet of current flows along $z$ at the domain edges, extending from $z=0$ to $-z_{\rm{spot}}$ and remaining uniform along $y$. The masked region ($z>0$) is modeled as an insulator with refractive index $n=3$, a value consistent with experimental observations\cite{Sebastian24}. We fit the amplitude of the edge currents to match the overall magnetic field amplitude predicted in the previous section, and then solve Maxwell’s equations for the space- and time-dependent magnetic field under the shielded region where the detector is placed, see Appendix~6.

Snapshots of the magnetic field distribution beneath the mask are shown at $ t = 0, 1, 1.5$ ps, respectively, in Fig.~\ref{fig:multi-bilayer3}(a), (b), (c). Furthermore, an animation of the propagation of the magnetic field under the mask can be found in Ref.~\citenum{duilio_2025_movie}. A propagating front $O(100)$-times larger than the equilibrium diamagnetic field amplitude, $|B_{\rm{ext}} \chi_d|$, is seen to reach the detector area. 
To further illustrate our point, we take an observation coordinate $x = 40 \; \rm{\mu m}$ (measured from the leftmost segment edge) and the following separations along $z$ from the pumped region: $\Delta z = 50 \; \mu$m, $\Delta z = 100 \; \mu$m, and $\Delta z = 150 \; \mu$m (corresponding, respectively, to the full, dashed, and dotted circles in Fig.~\ref{fig:multi-bilayer3}(c)). At these locations, we show in Fig.~\ref{fig:multi-bilayer3}(d) the time traces of the induced magnetic field. In agreement with experiment, we observe a pulse propagating along $z$ at the speed of light in the medium and progressively attenuating in strength.

To summarize, we now present a detailed comparison between the experimental observations of Ref.~\citenum{Sebastian24} and the predictions made within our theoretical framework:

a. \textit{Paramagnetic character of the response.} The peak magnetic field on the detector is paramagnetic and linear with applied external magnetic field. This is recovered by our theory. As discussed above, even though experimental parameters are deep in the nonlinear dynamical regime of the SG model, the response scales linearly with the magnetic field, see Fig.~\ref{fig:SGresults}(b);

b. \textit{Fluence dependence of the effect.} The induced magnetization measurements show a nonlinear behavior versus fluence, with a saturation trend at the highest fluences. We capture the fluence response seen in experiments for low fluences, see Fig.~\ref{fig:SGresults}(c). For higher fluences, we do not observe the reported saturation. To explain the saturating behavior, we point to the sensitive dependence of the phenomenon to dissipation demonstrated in Appendix~4. A heavy increase of dissipation, due to heating at such higher fluences, could indeed account for the observed saturation;

c. \textit{Propagation of the magnetic signal.} The magnetic pulse on the detector is paramagnetic and travels with the speed of light. We calculate the magnetic field distribution beneath the mask shielding the magnetic field detector, assuming the geometry and edge-current distribution illustrated in Fig.~\ref{fig:multi-bilayer3}(a). The medium beyond the mask's edge is modeled, at THz frequencies, using an effective dielectric constant matching with experimental observations\cite{Sebastian24}. This approach attributes the response observed in Ref.~\citenum{Sebastian24} to a giant dynamical paramagnetic response propagating through the region under the detector. As shown in Fig.~\ref{fig:multi-bilayer3}(d), increasing the propagation distance causes the signal to attenuate and shift to longer delays, consistent with the measurements in Ref.~\citenum{Sebastian24} and the behavior of a propagating electromagnetic wave;

d. \textit{Amplitude of the magnetic response.} The amplitude of the induced magnetic field is $B_{\rm{ind}} \approx \SI{6}{\micro T}$ for an external field of $B_{\rm{ext}} \approx \SI{10}{m T} $. Using realistic parameters, we obtain a result which quantitatively agrees with the experiment, $B_{\rm{ind}}/B_{\rm{ext}} \sim 10^{-3}$, see Fig.~\ref{fig:multi-bilayer2}(a). The intra-bilayer effective conductivity $\sigma_n$, which determines the damping coefficient as $\alpha = \frac{\sigma_n}{\epsilon_0 \omega_J}$, is hard to determine either from first principles or experimentally. Reducing (increasing) the latter parameter enhances (attenuates) the effect;

e. \textit{Temperature dependence of the effect.} While various parameters, such as $\omega_J$ and $\alpha$, can in principle contribute to the temperature dependence of the overall result for $B_{\rm{ind}}$, we start by focusing on the direct proportionality to the equilibrium diamagnetic field shown in Fig.~\ref{fig:multi-bilayer2}(b): $\left\langle B^{\rm{max}}_{\rm{ind}} (T) \right\rangle / B_{\rm{ext}} = - f(T) \, \chi_d (T)$, where $\left\langle ... \right\rangle$ indicates averaging over a segment of length $L$. The equilibrium diamagnetic response, $\chi_d$, is proportional to the local superfluid density, i.e., $\chi_d \propto n_{2D,s}(T)$. Assuming a preformed pair scenario, the critical temperature, $T_c$, is associated with a BKT transition, whereas the pseudogap temperature, $T^*$, is interpreted as that where Cooper pairs dissociate, thereby depleting the superfluid density. In the inset of Fig.~\ref{fig:multi-bilayer2}(b), we plot the temperature dependence of $\chi_d$, assuming a BCS form for the superfluid density: $n_{s}(T)/n_{s}(T_c) = \tanh \left( \sqrt{\kappa \frac{T^* - T}{T}} \right)$. The fact that the photo-induced effect, which is proportional to $\chi_d$, can be fitted using this expression, as shown in the experimental paper\cite{Sebastian24}, is highly suggestive for the nature of the pseudogap phase.
%To explain the experimental data, we derive the magnetic field distribution under the mask protecting the magnetic field detector, see Fig.~\ref{fig:multi-bilayer2}(a) and the Methods, by assuming that at THz frequency we can model the material under the mask by free Maxwell's equations with a refraction index $n \sim 3$ extracted from the experiment\cite{Sebastian24}. In this way, we attribute the response detected in Ref.~\citenum{Sebastian24} as the giant dynamical paramagnetism reaching the region underneath the detector.

%Recall that the amplified paramagnetic signal originates from instabilities of screening currents at edges. We argue that cracks and crystal defects, like the ones shown schematically in Fig.~\ref{fig:Sketch}(a), act as such edges. Since the typical distance between defects is $\rm 10 \, \mu m$, within a spot size $\rm 100 \, \mu m$, such edges will very likely be present. Using experimentally realistic parameters, we find that the induced paramagnetism is $O(100)$-times larger than the equilbrium diamagnetism, see Fig.~\ref{fig:multi-bilayer2}(b). Since $| \chi_d | \approx 10^{-5}$, this corresponds to $\frac{\langle B_{\rm{ind}} \rangle}{B_{\rm{ext}}} \sim 10^{-3}$, consistent with the experimental value for the maximum detected field. In Fig.~\ref{fig:multi-bilayer2}(c), we show that the effect is proportional to the equilibrium diamagnetism, which could explain the temperature dependence, see the Methods.

We emphasize that our mechanism is different from that proposed in Ref.~\citenum{Sebastian24}, where the paramagnetic signal is interpreted as originating from flux exclusion from a giant diamagnetic response under the pump. This mechanism can be distinguished from ours in future experiments that measure the magnetic field directly under the pumped regions, since opposite signs are predicted~\cite{Cavalleri_ongoing}. A second proposal is to place the mask so that it runs perpendicular to the layer orientation, instead of parallel as in Ref. \citenum{Sebastian24}. This geometry is close to that treated in Sec. 4, if we interpret the sample between $0$ and $L$ as the pumped region and the area under the mask as the region outside one edge of the sample. In this case, we expect the emitted magnetic flux to be bimodal and mostly canceled upon a short time integration, as seen in Fig.~\ref{fig:emittedfield}, implying a much smaller observed magnetic field. 

\section{6. Flux-Floquet instability in other $U(1)$ systems}

\begin{figure}
    \centering
    \includegraphics[scale=1.2]{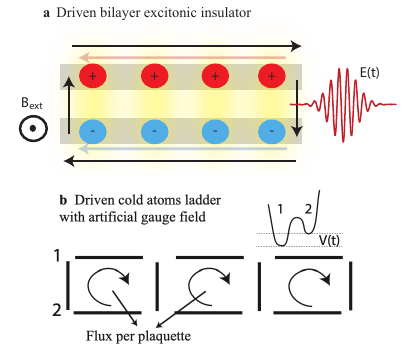}
    \caption{\textbf{Flux-Floquet instability in other systems.} \textbf{a} Excitonic insulators correspond to condensates of electron-hole pairs. In bilayer excitonic insulators, electrons (negatively charged blue spheres) and holes (positively charged red spheres) reside in different layers but are nevertheless bound by the Coulomb interaction. Such systems are expected to exhibit diamagnetic screening currents through the superflow of excitons in the presence of a magnetic field, a phenomenon referred to as the ``excitonic'' Meissner effect\cite{Yuelin26}. This effect is analogous to the counterflow diamagnetic screening in bilayer Josephson junctions. The physical diamagnetic currents are depicted by black arrows, while the motion of charge in each layer is indicated by the blue and red arrows, respectively. This effect is typically too small to be observed directly, but it could become observable through the giant dynamical paramagnetism induced under driving by an electric field polarized perpendicular to the layers, as sketched. \textbf{b} Sketch of a cold-atom ladder with artificial gauge fields. The links correspond to the bonds between sites on the two-legged ladder. Driving of these ladders in the presence of an artificial gauge field can be achieved by applying a time-dependent potential difference $V(t)$ between the two legs of the ladder. }
    \label{fig:otherSystems}
\end{figure}

Our primary motivation in this manuscript is to provide a theoretical understanding of the giant magneto-optic effect observed in driven high-$T_c$ cuprates\cite{Sebastian24}. However, the flux-Floquet instability identified in this work is a universal phenomenon that should manifest in a variety of solid-state and ultracold atomic systems. In this section, we discuss several relevant platforms, see Fig.~\ref{fig:otherSystems}.

%\textbf{c} Sketch of driven fluctuating superconductors above $T_c$. The gray ellipsoids represent local superconducting puddles in a fluctuating superconductor. The links between puddles represent Josephson couplings. The presence of local superconducting correlations, as depicted here, gives rise to weak diamagnetic screening current loops flowing across effective local Josephson-junction networks, indicated by black arrows. In the presence of strong external driving, Josephson currents aligned along the driving direction can trigger the flux-Floquet instability, giving rise to large paramagnetic currents depicted by red arrows.

\subsection{6.1 Extended Josephson devices}

Extended Josephson systems are superconducting weak links whose spatial extent is comparable to or greater than the Josephson penetration depth (defined above)~\cite{Likharev86}. The dynamics of extended Josephson junctions in the presence of magnetic field is also described by the SG-type model in Eqs.~\eqref{eq:drivenEOM}--\eqref{eq:magnStatic}. One important difference is the value of parameter $\beta$ controlling the equilibrium screening currents. In the system considered in this paper, it is very small, $\beta \sim 10^{-5}$. In an Al-based nanodevice with typical inter-electrode distance $d \sim 1~\mathrm{nm}$, typical electrode thickness $t_e \sim 50~\mathrm{nm}$ and plasma frequency $\omega_{pl} \sim 2 \pi \times 10^3~\mathrm{THz}$, we find $\beta_{Al} = \frac{t_e \omega_{pl}^2 d}{2 c^2}\sim 10^{-1}$. This much larger value of $\beta$ markedly favors the overall magnitude of the instability-induced enhancement. Realistic Josephson critical current densities of $J_c \sim 10^6~\mathrm{A/m^2}$, see Ref.~\citenum{Abdumalikov2006}, give a Josephson plasma frequency of $\omega_1 \sim 2 \pi \times 100~\mathrm{GHz}$. This yields an estimate of order $1~\mathrm{mV}$ for the critical AC voltage required to trigger the instability, applied at frequencies around $10\text{–}100~\mathrm{GHz}$. Such microwave AC drives are available in the existing literature on extended Josephson junctions~\cite{Barkov2004}.

The flux-Floquet instability can be used to dramatically amplify the recently reported the superconducting diode effect of extended Josephson junctions in the presence of magnetic field~\cite{Sundaresh23, Guarcello2024efficiency}. Notably, the paramagnetic character of the instability yields a diode response with a sign opposite to that observed in the absence of driving. We plan to present this analysis in a subsequent publication~\cite{DeSantis_ongoing}. Furthermore, the flux-Floquet instability in extended Josephson junctions can be used to increase sensitivity to magnetic fields, which has implications for quantum information processing, including qubit readout protocols~\cite{Osborn2020reversible, Cai2024fluxon, Grankin_2024, Wustmann2025simulation, Walsh2021josephson, Pankratov2025detection}.

\subsection{6.2 Excitonic insulators}

Excitonic insulators provide another example of a broken-$U(1)$ system, in which the condensate consists not of Cooper pairs but of electron-hole pairs, namely excitons, bound by the strong Coulomb interaction. Two-dimensional van der Waals (vdW) materials have opened the way to realizing bilayer excitonic insulators\cite{Nguyen26}, in which positively charged holes and negatively charged electrons are spatially separated into two different atomically thin materials by a thin insulating layer such as hBN, see Fig.~\ref{fig:otherSystems}(a). A key open question in this field is how to establish the existence of superfluid transport\cite{Baldini23}. Experiments on bilayer excitonic-insulator candidates, such as MoSe2/WSe2 heterostructures, have so far demonstrated perfect Coulomb drag\cite{Nguyen25}. However, while this is a necessary condition, Coulomb drag alone does not establish the presence of dissipationless exciton flow. Such a flow can be probed more directly by applying an in-plane magnetic field and detecting an ``excitonic'' Meissner effect\cite{Yuelin26}, which would give rise to a diamagnetic contribution analogous to the diamagnetic response discussed in this paper for a single-bilayer SC.

The diamagnetic response of an excitonic insulator in the presence of an in-plane magnetic field has previously been addressed in the literature in the context of excitonic insulators in quantum Hall bilayers\cite{Fogler01,Yang96}. These works show that, in the presence of weak tunneling between the two layers, the equations of motion are governed by the same form as those of a single-bilayer extended Josephson junction, see Eqs.~\eqref{eq:drivenEOM}--\eqref{eq:magnStatic}.

While this effect is crucial for establishing the existence of in-plane superfluidity in excitonic insulators, it has not been experimentally observed to date. In the context of vdW materials, the reason is that, using the formulas derived in this paper for counterflow diamagnetism, the expected signal is too small\footnote{The exciton-induced diamagnetism is proportional to $\beta_{ex} = \frac{e^2 n_{ex,2D} d_{\rm hBN}}{2 m_{ex} \epsilon_0 c^2}$. For a typical material\cite{Gunda_25} with $n_{2D} \sim 10^{12} \rm{cm}^{-2}$ and taking $m_{ex} \sim \mathcal{O}(1)\, m_{el}$, and assuming an hBN thickness of $2\,\rm nm$, we find $\beta_{ex} = - \chi \sim 10^{-6}$.}: $\beta_{ex} = - \chi \sim 10^{-6}$.

The flux-Floquet instability uncovered in this article can be exploited in these systems to amplify the magnetic signal, making it experimentally detectable in the presence of driving. At the same time, vdW heterostructures provide a highly tunable platform for realizing the dynamical soliton injection described in Sec.~3. 

As in the previous section, the precise AC voltage required to trigger the instability depends on the device parameters, most notably the Josephson critical current density. As an estimate, we consider Ref.~\citenum{Ziyuan21}, which reports a typical value $J_c \sim 10^9~\mathrm{A/m^2}$ for an electron--hole excitonic bilayer. For an interlayer separation of $d = 1~\mathrm{nm}$, this corresponds to a threshold voltage of approximately $V_{\mathrm{crit}} \sim 10~\mathrm{mV}$ across the electrodes, or equivalently to a critical electric field $E_{\mathrm{crit}} \sim 10~\mathrm{MV/m}$.

\subsection{6.3 Ultracold atomic systems}

The SG model widely emerges as a low-energy effective theory in bosonized low-dimensional quantum systems~\cite{Giamarchi} and can therefore be readily engineered in cold-atom platforms. In particular, the continuous case of two Raman-coupled condensates in the presence of external flux can be mapped onto the SG model, as discussed in Refs.~\citenum{Gritsev07,Kasper20,Lovas22}. Two-legged bosonic ladders in the presence of synthetic gauge fields\cite{Wybo23,Impertro24,Impertro25} likewise fall within this broader class of systems, see Fig.~\ref{fig:otherSystems}(b) for a sketch. For two coupled condensates, the counterpart of the strong AC optical drive is a time-dependent modulation of the potential-energy difference between the two one-dimensional wires~\cite{Schweigler2017,Schweigler2021}. Moreover, ultracold-atom experiments provide access to the quantum regime of these systems, making it especially interesting to explore how the flux-Floquet instability is modified beyond the semiclassical regime considered in the present paper. A major advantage of studying this instability in cold-atom systems is the ability to image spatially resolved currents. Ref.~\citenum{Impertro25} has already demonstrated this capability for equilibrium counterflow currents. This platform therefore provides a promising route toward experimentally testing our spatially resolved predictions for the pre-soliton regime.

\section{Discussion}

It is useful to position our work in the general landscape of ultrafast probes of electron states and photo-induced phases. A ubiquitous challenge in this field is identifying the correct methodology for analyzing experimentally measured properties of transient states. A common approach is to use the quasi-static perspective, in which experimental signatures are interpreted in direct analogy with equilibrium systems. This approach demonstrably breaks down when transient regimes feature qualitatively different properties from the static systems. In these cases, one needs to take into account the dynamical character of photoinduced states. Notable examples include light induced renormalization of electronic bands, which can lead to Floquet topological systems \cite{McIver2020,choi2024} or changes in terahertz optical properties due to photoexcited collective modes, that result in new features in reflectivity and even light amplification \cite{Shimano2020,vonHoegen22,Haque2024,Dai2021} (for related studies, see also analysis of photonic time crystals \cite{Lustig2023}). The theoretical analysis presented in this paper suggests that experimental results of the Hamburg group \cite{Sebastian24} should be interpreted from this dynamical perspective rather than by analogy with the equilibrium Meissner effect. We point out that in the case of bilayer YBCO, there is a large separation in Josephson coupling strengths within bilayers and between them. This gives rise to an extended temperature range, in which bulk diamagnetic response is suppressed because there can be no static flow of screening currents between the bilayers; however, charge dynamics within bilayers is dominated by preformed Cooper pairs. Thus, we argue that experimentally observed light-induced giant paramagnetic response in YBCO is a result of the flux-Floquet instability of strongly driven Josephson plasmons within individual bilayers in the presence of static magnetic field. We demonstrate that this phenomenon reveals itself in the macroscopic properties of the system, because at finite frequencies Maxwell’s displacement currents can supplant the charge currents. The phenomenon of giant paramagnetic response triggered by dynamic instability has no static analogues. Our work demonstrates that in layered SCs above $T_c$, short-range superconducting correlations can be revealed in the far out of equilibrium dynamics, even when the canonical linear response probes, such as diamagnetic susceptibility, do not provide indications of enhanced pairing fluctuations. Hence, the photomagnetic phenomena observed in Hamburg experiments provide another demonstration of the importance dynamical aspects of photoinduced transient states in ultrafast experiments in solids.

In regards to the physics of the pseudogap phase, the phenomena uncovered in this paper crucially depend on the existence of local pairing correlations. While evidence for such correlations has been reported up to a temperature of $180 \, \rm K$ \cite{dubroka2011evidence,TajimaPRL}, the same cannot be stated up to $T^* \approx 300-400\, \rm K$. If our interpretation is correct, the magnetic field response observed in experiments\cite{Sebastian24} would represent the first evidence of local superconducting correlations surviving up to the pseudogap temperature $T^*$. This point of view is further supported in a separate paper \cite{michael2025parametrically}, which shows that the same model can explain the optical response and second harmonic generation experiments \cite{vonHoegen22,taherian24}. The origin of the pseudogap has been under debate for decades. Much of the discussions has centered on anti-ferromagnetic spin fluctuations, identifying the gap as a spin gap due to some underlying resonating valence bond (RVB) scenario\cite{lee2006doping}, or suppression of spectral weight due to strong correlation near the Mott transition\cite{wu2018pseudogap}. Occasionally, pairing scenarios such as pair density waves have also been proposed\cite{lee2014amperean,agterberg2020physics}. It is worth noting that in the RVB scenario, the pseudogap phase arises from spinon pairing or its gauge equivalence from which $d$-wave SC emerges, so that between $T_c$ and $T^*$ short-range order in the local pair phase is expected\cite{kotliar1988superexchange,lee19982,senthil2009coherence,christos2023model, christos2024emergence}. We also point out that recent low-temperature local shot noise measurements using scanning tunneling microscopy (STM) have been used to infer that the large energy gap ($40$ to $60$ meV) is a pairing gap~\cite{niu2024equivalence}. This aligns with the picture presented in this paper, as it is reasonable to expect that such a large gap would give rise to local pairing persisting to high temperatures. We hasten to add that the existence of local pairing does not necessarily imply that the pairing is responsible for the formation of the pseudogap. Nevertheless, the notion of local pairing surviving up to $T^*$ will clearly have a strong impact towards resolving the pseudogap puzzle.

%The phenomenon we observe crucially depends on the existence of local pairing correlations. While evidence for such correlations has been reported up to a temperature of $180$ K\cite{dubroka2011evidence}, the same cannot be stated up to $T^*$. If our interpretation is correct, the magnetic field response observed in experiments\cite{Sebastian24} would represent the first evidence of local superconducting correlations surviving up to the pseudogap temperature $T^*$. 

%At its core, our theory predicts a paramagnetic response, while the authors of the Hamburg experimental work\cite{Sebastian24} have interpreted their data as originating from a diamagnetic response. This disagreement can be resolved in future experiments by adopting a geometry where a thin film of YBCO is pumped from above, with the magnetic field detector placed beneath the sample\cite{Cavalleri_ongoing}. In such a geometry, our mechanism would still predict a paramagnetic detected field.

\section{Methods}
\label{sec:methods}

\paragraph{Electrodynamics of the counterflow pseudogap state.} In the pseudogap, there is no long-range superconducting coherence, so both the in-plane and out-of-plane DC currents are dissipative due to the proliferation of vortices. However, strong Josephson coupling within a single bilayer leads to locking of vortex lines passing through it. This gives rise to intrabilayer superconducting coherence\cite{fertig2002deconfinement} and dissipationless in-plane currents flowing in opposite directions\cite{Homann24}. The minimal model capturing these basic ingredients can be formulated as follows for the currents:
\begin{align}
    \partial_t j_{z,1} =& \, \frac{e^* d_1}{\hbar} J_{c,1} E_{z,1} \cos \theta_1, \label{eq:jz1} \\
    \partial_t j_{x,1} =& \, \frac{(e^*)^2 n_s}{m} E_{x,1} + \gamma_d \left( j_{x,1} + j_{x,2} \right), \label{eq:jx1} \\
    \partial_t j_{x,2} =& \, \frac{(e^*)^2 n_s}{m} E_{x,2} + \gamma_d \left( j_{x,1} + j_{x,2} \right), \label{eq:jx2}
\end{align}
for a generic unit cell, where the label `$z,1$' (`$z,2$') denotes the average of a $z$-component between layers separated by a distance $d_1$ ($d_2$), and `$x,1$' (`$x,2$') an $x$-component evaluated at the $z = z_1$ ($z = z_2$) layer. Within the bilayer, the superconducting coherence is encoded in the Josephson relation. We assume interbilayer coherence to be completely lost, in accordance with the experimental and numerical observation of the lower Josephson plasmon's disappearance at $T = T_c$ in YBCO. Furthermore, the total in-plane current, $j_{x,1} + j_{x,2}$, is dissipative due to vortex motion\cite{Nelson79} ($\gamma_d$ is the corresponding friction coefficient), while the counterflow current, $j_{x,1} - j_{x,2}$, remains dissipationless. This assumption can  be relaxed by adopting a two-fluid model, where a normal component is added to the counterflow current. This replaces $(e*)^2 n_s/m$ by $(e^*)^2 n_s / m +i \omega \sigma_{2D}$ and introduces an imaginary part to the parameter $\beta$, where $\sigma_{2D}$ is the in-plane 2D normal conductivity. As a result, an extra dissipative term of the form $\partial_t \partial_{xx}$ is added to the equation below. We find that this term has very little effect on the results, due to the extra $x$ derivatives.

As we describe in Appendix~1, combining Eqs.~\eqref{eq:jz1}--\eqref{eq:jx2} with a normal-fluid dissipative current contribution and Maxwell's equations leads to the following effective electrodynamic equations for the momentum $q_z = 0$ response in the pseudogap phase:
\begin{widetext}
\begin{align}
    \partial_t^2 \theta_1 + \frac{\sigma_{n,1}}{\epsilon_0} \partial_t  \theta_1 - \frac{c^2 (d_1+ d_2 \beta)}{d_1 + d_2(1 + \beta)}  \partial_x^2 \theta_1 + \omega^2_{J,1} \sin \theta_1 =& \, \frac{e^* d_1}{\hbar} \partial_t E_p(t) + \frac{c^2 d_1}{d_1 + d_2(1 + \beta)} \partial_x^2 \theta_2, \label{eq:DrivenMulti1} \\
    \partial_t^2 \theta_2 + \frac{\sigma_{n,2}}{\epsilon_0}\partial_t \theta_2 - \frac{c^2 d_2 (1 + \beta)}{d_1 + d_2(1 + \beta)} \partial_x^2 \theta_2 =& \, \frac{e^* d_2}{\hbar} \partial_t E_p(t) + \frac{c^2 d_2}{d_1 + d_2(1 + \beta)} \partial_x^2 \theta_1, \label{eq:DrivenMulti2} %- \kappa_2 \partial_t \theta_2
\end{align}
\end{widetext}
which describe a set of capacitively coupled long Josephson junctions. Here $\sigma_{n,1}$ denotes the normal c-axis conductivity between members of the bilayer and leads to dissipation. Formally, we introduce a similar quantity $\sigma_{n,2}$ in Eq. \eqref{eq:DrivenMulti2} to describe dissipation of $\theta_2$. However, due to the weak tunnel coupling between bilayers, the actual normal conductivity is negligible. The $\sigma_{n,2}$ originates from the fact that the cavity for the radiation field formed by the region between bilayers is leaky. The cavity Q leads to dissipation of photon modes, which can be parameterized by $\sigma_{n,2}$. Similar to the single bilayer case, we introduce the dimensionless intra-bilayer and inter-bilayer damping coefficients $\alpha_1 = \frac{\sigma_{n,1}}{\epsilon_0 \omega_{J,1}}$ and  $\alpha_2 = \frac{\sigma_{n,2}}{\epsilon_0 \omega_{J,1}}$, respectively. Since $\alpha_1$ and $\alpha_2$ have different physical origins, they can take on very different values, as chosen in Fig. \ref{fig:multi-bilayer} in the main text. The main approximation made here is to work with averaged quantities between different layers. We note in passing that the perturbed SG model, see Eq.~\eqref{eq:drivenEOM} in the main text, is readily recovered in the $d_2 \to \infty$ (single bilayer) limit of Eq.~\eqref{eq:DrivenMulti1}.

As shown in Appendix~1, the gauge invariant phase differences between layers, $\theta_1$ and $\theta_2$, can be directly related to the average out-of-plane electric fields, $E_{z,1}$ and $E_{z,2}$:
\begin{align} 
\partial_t \theta_1 =& \, \frac{e^* d_1}{\hbar} E_{z,1}, \\
\partial_t \theta_2 =& \, \frac{e^* d_2}{\hbar} E_{z,2}.
\end{align}
By relating the discontinuity of $B$ across each layer to the in-plane currents, simple expressions for $\theta_1$ and $\theta_2$ in terms of the average in-plane magnetic fields between layers, $B_{y,1}$ and $B_{y,2}$, also hold:
\begin{align}
    \partial_x \theta_1 =& \, \frac{e^* d_1}{\hbar} \left( B_{y,1} + \frac{B_{y,1} - B_{y,2}}{\beta} \right), \\
    \partial_x \theta_2 =& \, \frac{e^* d_2}{\hbar} \left( B_{y,2} + \frac{B_{y,2} - B_{y,1}}{\frac{d_2}{d_1} \beta} \right) .
\end{align}

For earlier theoretical studies of Josephson plasmons in layered SCs below $T_c$, see Refs.~\citenum{Bulaevskii94,Kleiner94,vdMarel01,Koyama01,Gabriele21,Gabriele22,Sellati23}.

\paragraph{Diamagnetism due to the dissipationless counterflow state.} To find the static response to an external magnetic field $B_{\rm{ext}}$, we consider a semi-infinite geometry with a boundary to air at $x=0$, and impose the BC: $ \left. B_{y,1} \right|_{x=0} = \left. B_{y,2} \right|_{x=0} = B_{\rm{ext}} $. In Appendix~2, the solution of Eqs.~\eqref{eq:DrivenMulti1}--\eqref{eq:DrivenMulti2} is shown to be
\begin{align}
    B_{y,1}(x) =& \, \left( \frac{\beta}{1 + \beta} e^{-x/\lambda} + \frac{1}{1 + \beta} \right) B_{\rm{ext}}, \\
    B_{y,2}(x) =& \, B_{\rm{ext}},
\end{align}
where $\lambda = \sqrt{\frac{\beta}{1+ \beta}} \frac{c}{\omega_J} $ is the bilayer penetration depth. For $x \gg \lambda$, we find a screened intrabilayer magnetic field, $B_{\rm{sc}} = \frac{B_{\rm{ext}}}{1 + \beta} $, whereas there is no interbilayer screening. Averaged over one unit cell, the effective diamagnetic susceptibility is given by:
\begin{equation}
    \chi_d = - \frac{d_1 \beta}{D} = -\frac{d_1 (e^*)^2 n_s}{D \, 2 \epsilon_0 m c^2},
\end{equation}
where, in the second step, we expressed $\beta$ in terms of microscopic parameters of the counterflow state, such as the 2D superfluid density, $n_s$, in each layer.

From existing data in the pseudogap phase of YBCO\cite{Cooper12}, a geometric analysis of magnetic susceptibility found an enhanced diamagnetic contribution for the in-plane susceptibility, which is sensitive to counterflow currents, versus the out-of-plane magnetic susceptibility, with $\delta \chi \sim 10^{-5}$. Attributing this difference to the counterflow state, we estimate $\beta \sim 10^{-5}$. 

\begin{figure}
    \centering
    \includegraphics[scale = 1]{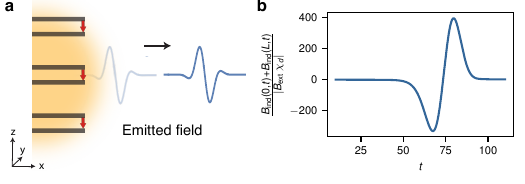}
    \caption{ \textbf{Magnetic field emitted from pumped YBCO.} \textbf{a} Sketch of the magnetic pulse emitted from the right edge ($x > L$) of a YBCO region illuminated by the pump field. An analogous emission process occurs at the left edge ($x < 0$) of the segment (not shown). \textbf{b} Average induced magnetic field at the edges,  $B_{\rm{ind}}(0, t) + B_{\rm{ind}}(L, t)$, normalized by the equilibrium diamagnetic field amplitude, $|B_{\rm{ext}} \chi_d|$. We observe a bimodal emission profile. The chosen simulation parameters are identical to that of Fig.~\ref{fig:multi-bilayer}, see the main text, and time is rescaled to $1/\omega_{J,1}$. }
    \label{fig:emittedfield}
\end{figure}
\paragraph{Radiative BC.} Focusing on the physics between two cracks, BC are required to model the electrodynamics outside the pumped segment. Specifically, as localized paramagnetic currents are generated and amplified in the pre-soliton regime, an equal and opposite flux is expected to be emitted from the segment's edges, such that the total flux is conserved. By imposing appropriate continuity conditions for the average electric and magnetic fields at the segment-to-exterior interface, as illustrated in Appendix~1, we obtain:
\begin{align}
    \left. \left[ \partial_x \theta'_1 + \partial_x \theta'_2 \mp \frac{1}{c} \left( \partial_t \theta'_1 + \partial_t \theta'_2 \right) \right] \right|_{x=0,L} =& \, \frac{e^* D}{\hbar} B_{\rm{ext}}, \\ 
    \left. \left( \frac{\partial_x \theta'_1}{d_1} - \frac{\partial_x \theta'_2}{d_2} \right) \right|_{x=0,L} =& \, 0,
\end{align}
%\frac{n}{c}
%, and $n$ is the refractive index
where the $-$ ($+$) sign applies to the $x = 0$ ($x = L$) boundary. As motivated above, we decomposed our dynamics as $\theta_{1,2}(x,t) = \tilde{\theta}_{1,2}(t) + \theta'_{1,2}(x,t)$, with $\tilde{\theta}_{1,2}$ being the homogeneous parts, strongly driven and evolving over the pump period's timescale, and $\theta'_{1,2}$ mostly keeping track of slower (over the pump envelope's timescale) and spatially localized features.

In the segment's exterior, concentrating on the right-propagating mode sketched in Fig.~\ref{fig:emittedfield}(a), we then have that
\begin{equation}
    B_{\rm{em}}(x > L, t) = B_{\rm{ind}}(L, t - (x - L)/c), %n(x - L)/c),
    \label{eq:Bem}
\end{equation}
where $B_{\rm{ind}} = \frac{\hbar}{e^* D} (\partial_x \theta'_{1,\rm{ind}} + \partial_x \theta'_{2,{\rm{ind}}})$ is the $z$-averaged field induced within a unit cell. Taking also into account the left-propagating mode for $x < 0$, total flux conservation readily implies that
\begin{equation}
    B_{\rm{ind}}(0, t) + B_{\rm{ind}}(L, t) = -\frac{1}{cD} \partial_t \Phi_{\rm{ind}},
    \label{eq:fluxcons}
\end{equation}
%\frac{n}{cD}
with $\Phi_{\rm{ind}}$ being the induced flux inside the segment. Therefore, the information about the emitted fields is encoded in the $B_{\rm{ind}}(0, t) + B_{\rm{ind}}(L, t)$ time traces, shown in Fig.~\ref{fig:emittedfield}(b) for our representative multi-bilayer scenario. The emission profile is bimodal, which can be understood as a consequence of flux conservation. To conserve total flux, as the induced flux in the segment increases, a diamagnetic pulse is emitted outward from the segment. Conversely, when the induced flux decreases, a paramagnetic flux is emitted. In general, we expect any short-lived paramagnetic or diamagnetic response to give rise to a bimodal emission, which averages to zero over time, in accordance with Eq.~\eqref{eq:fluxcons}. %This observation further supports the conclusion that the detected signal\cite{Sebastian24} is the direct magnetic response of the sample under the detector, rather than emitted radiation. 

\begin{figure}
    \centering
    \includegraphics[scale = 1]{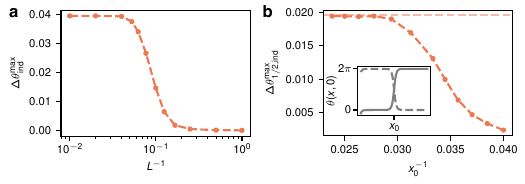}
    \caption{ \textbf{Size analysis and impact of solitonic initial conditions.} \textbf{a} Dependence of the maximum photo-induced phase difference versus the (normalized) inverse length of the bilayer segment. If $L < 10$, i.e., if the segment is smaller than the characteristic length of the pre-soliton shown in Fig.~\ref{fig:SGoverview}, the effect is heavily suppressed, whereas for $L < 10$ the response saturates, demonstrating that this is indeed a boundary effect. \textbf{b} Half-size maximum photo-induced phase difference, averaged over soliton and antisoliton initial conditions (represented in the inset), versus the inverse coordinate $x_0^{-1}$. Here the segment length is $L = 200$. Saturation to the soliton-free response (see the dashed horizontal line) is observed for $x_0^{-1} \lesssim 0.03$, whereas the response gradually decays for larger $x_0^{-1}$. The following (normalized) parameters are used in both plots: $V_0 = 3.5$, $q = 1.4 \times 10^{-4}$, $\alpha=0.5$, and $\beta = 10^{-5}$.}
    \label{fig:SGmethods}
\end{figure}
\paragraph{SG pre-soliton mechanism: length dependence and robustness to thermal (anti)solitons.} Being a spatially extended excitation, a pre-soliton can only form if the bilayer system is large enough to host it. To this end, we explore the SG model's ($\theta_1 \equiv \theta$ here) response versus the (normalized) inverse segment length $ 0.01 \leq L^{-1} \leq 1 $, see Fig.~\ref{fig:SGmethods}(a). We observe a saturating behavior, with a characteristic inverse length of $\sim 0.1$, which is consistent with the pre-soliton size of $\sim 20$ we deduce from, e.g., Fig.~\ref{fig:SGoverview}(d) and (e). The latter point confirms, at the SG level, that this is indeed a boundary effect.

We have not explicitly studied the influence of thermally excited solitons and antisolitons on our mechanism. For $T \gtrsim T_c$, our choice is justified because the density of such fluctuations is presumably very small by virtue of the large value of the intrabilayer Josephson coupling. However, at higher temperatures, this is no longer a valid assumption. We can estimate the typical distance between two solitons through a high temperature expansion~\cite{Koshelev1996} to be given by in-plane coherence length, $\xi$. 

For the single bilayer, we investigate the robustness of the light-induced paramagnetism phenomenon to (anti)solitons by repeating the SG calculations for $V_0 = 3.5$ and $q = 1.4 \times 10^{-4}$ using the initial conditions (in dimensionless form):
\begin{equation}
\theta(x,0) = - q \frac{e^{-x} - e^{-(L - x)}}{1 + e^{-L}} + 4 \arctan e^{\pm (x - x_0)},
\label{eq:icsoliton}
\end{equation}
where the first terms encodes the static diamagnetic response within the segment, and the second term corresponds to a SG soliton ($+$) or antisoliton ($-$) at a distance $x_0$ from the $x = 0$ edge.

We work with $0 < x_0 < L/2$, and evaluate the system response in terms of the half-size maximum photo-induced phase difference, i.e., $\Delta \theta^{\rm{max}}_{1/2,\rm{ind}} = \mbox{max} \{ \left.\theta\right|_{x=0}^{x=L/2} - \left. \theta_{\rm{eq}} \right|_{x=0}^{x=L/2} \}$. In particular, expecting the same average population of solitons and antisolitons, in Fig.~\ref{fig:SGmethods}(b) we plot $\Delta \theta^{\rm{max}}_{1/2,\rm{ind}}$ versus $x_0^{-1}$ upon averaging over the two ($\pm$) different solitonic initial conditions, see also the sketch in the inset. Our mechanism shows robustness to the presence of (anti)solitons, as the response saturates to the soliton-free counterpart (see the dashed horizontal line) for $x_0^{-1} \sim 0.03 $. Beyond this threshold, we observe a gradual suppression of the effect. In fact, when (anti)solitons are very close to the edge, they can interfere with the pre-soliton emergence, as well as scatter from the boundary, thereby leading to somewhat convoluted nonlinear transients.

\begin{figure}
    \centering
    \includegraphics[scale = 1]{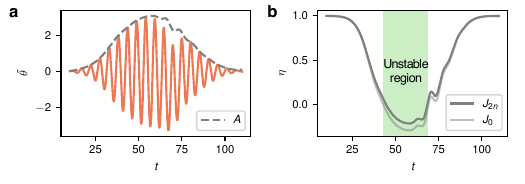}
    \caption{\textbf{Homogeneous solution and parametric drive.} \textbf{a} $\tilde{\theta}$ trajectory (full line), and its smoothly varying envelope $A$ (dashed line), versus time. \textbf{b} Parametric drive $\eta(A(t))$, comparing the $J_0$-only result, see the light gray line, with that accounting for higher-frequency ($J_{2n}$) corrections, see the gray line. The green shaded area denotes the unstable, $\eta < 0$, region. The (normalized) parameters values $V_0 = 3.5$ and $\alpha=0.5$ are used in both plots.}
    \label{fig:Floquetmethods}
\end{figure}
\paragraph{Floquet framework.} According to Eqs.~\eqref{eq:pendulum}--\eqref{eq:effectiveBC}, the homogeneous solution $\tilde{\theta}$ follows the driven damped pendulum equation, and its slowly varying amplitude $A(t)$ enters the linearized dynamical equation for the inhomogeneous part $\theta'$ as a parametric drive $\eta(A)$. Specifically, as shown in Appendix~3, a Floquet expansion yields the expression:
\begin{equation}
    \eta(A) = J_0(A) + 2 \sum_n \frac{ J_{2n}(A) J_{-2n}(A) \left[ (2n \omega_p)^2 - J_0(A) \right] }{\left[ (2n \omega_p)^2 - J_0(A) \right]^2 + \alpha^2 (2 n \omega_p)^2 } ,
\label{eq:Jeff}
\end{equation}
where $J_i$ is the $i$-th Bessel function of the first kind.

We numerically solve for the $\tilde{\theta}$ trajectory, and subsequently extract a smoothly varying envelope $A(t)$ via a low-pass-filtered Hilbert transform, see Fig.~\ref{fig:Floquetmethods}(a). We are then in position to evaluate the parametric drive $\eta(A(t))$, which features a leading contribution, see the `$J_0$' term in Eq.~\eqref{eq:Jeff}, as well as higher-frequency corrections, see the `$J_{2n}$' summation over $n$ in Eq.~\eqref{eq:Jeff}. In this regard, in Fig.~\ref{fig:Floquetmethods}(b) we display the parametric drive relative to the time trace of Fig.~\ref{fig:Floquetmethods}(a), comparing the result coming from the leading contribution alone, see the light gray line, with that obtained by including the higher-frequency corrections for $n \leq 5$, see the gray line. The two plotted $\eta$ curves qualitatively look similar: their equilibrium value is one, and they cross zero---reaching negative values---due to the strongly nonlinear drive, see the green shaded area in Fig.~\ref{fig:Floquetmethods}(b). The $\eta < 0$ region is responsible for the unstable behavior in the slow variable $\theta'$, and therefore we expect the system's response to be sensitive to the appreciable quantitative difference between the two results shown in Fig.~\ref{fig:Floquetmethods}(b).

We find that including higher-frequency corrections tames the instability by a noticeable amount (as compared to the $J_0$-only result), bringing the Floquet theory into remarkably close agreement with the SG prediction, as discussed in the main text. As a side note, we observe that only the very first terms in the `$J_{2n}$' summation are seen to matter in the present scenario, which motivates our choice of evaluating the $J_{2n}$ sum up to $n = 5$.

\paragraph{Numerical techniques.} To computationally handle the PDEs (in general, systems of coupled PDEs) presented in this article, we formulate them as systems of coupled first-order ODEs by employing second-order finite-difference approximations for the space derivatives. For instance, considering the (single bilayer) SG case for the field $\theta(x,t)$, we discretize the spatial domain as $N$ points separated by steps of length $\Delta x$, and we define the restriction $\theta(n \Delta x, t) \equiv \theta_n (t)$ for $n = 1, ..., N$. We then obtain a set coupled first-order ODEs for the quantities $ \partial_t \theta_n \equiv v_n (t)$ and $ \partial_t v_n \equiv a_n(t) $, which can be solved via standard numerical routines, such as \texttt{scipy.integrate}'s \texttt{odeint}, for a discrete set of $M$ times separated by intervals of duration $\Delta t$.

For the single bilayer numerical runs, we typically use the following discretization steps values: $\Delta x \approx 0.05$ and $\Delta t \approx \Delta x / 5$. We tested the reliability of our numerical findings in multiple ways: (i) we checked the stability and consistency of the simulation outcomes upon systematic variation of the $\Delta x$ and $\Delta t$ discretization steps; (ii) we performed extensive preliminary runs via two other independent numerical approaches, the first being an implicit implementation based on Thomas' algorithm and the other making use of \texttt{Mathematica}'s \texttt{NDSolve}, and we observed close agreement between the different methods on the discussed phenomena; (iii) within the pre-soliton regime, we reproduced the relevant features of the output of the full nonlinear (SG) model by recasting the problem in the Floquet picture, as addressed above. Similar assessments were conducted also for the multi-bilayer runs, which are presented in this article only for the Floquet effective theory, using step sizes of $\Delta x \approx 0.1$ and $\Delta t \lesssim \sqrt{\beta} \Delta x / 2$, the latter choice being determined by stability considerations on the linearized model.

As a concluding technical remark, we note that throughout the paper, results are presented after convolution with a Gaussian profile over several driving periods. This reflects the fact that the experiments\cite{Sebastian24} originally motivating our work use detectors that average over that timescale.

\section{Data and Code Availability}

The data presented in this work is produced directly by the codes provided below. The data can also be provided upon request. 

An example of the code used to produce the data in the paper is openly available on GitHub, see Ref.~\citenum{duilio_2024}.

\section{Acknowledgements}

We acknowledge fruitful discussion with Andrea Cavalleri, Angelo Carollo, Hope Bretscher, Immanuel Bloch, Michele Buzzi, Pavel Dolgirev, Monika Aidelsburger, Sebastian Fava, Giovanni De Vecchi, Claudio Guarcello, Gregor Jotzu, Yuli Lyanda-Geller, Andrew Millis, Walter Metzner, Alex Potts, Gil Refael, Leonid Rokhinson, Subir Sachdev, Bernardo Spagnolo, Davide Valenti, and Jukka Vayrynen. M.H.M. is grateful for the financial support received from the Alex von Humboldt postdoctoral fellowship. D.D.S. acknowledges the support of the Italian Ministry of University and Research (MUR) and ETH Zurich for the hospitality. E.A.D. acknowledges support from ETHZ, the SNSF project 200021\textunderscore212899, the ARO grant number W911NF-21-1-0184, and NCCR SPIN (SNSF grant number 225153). P.A.L. acknowledges support from DOE (USA) office of Basic Sciences Grant No. DE-FG02-03ER46076.

% \section{Author Contribution}

% M.H.M., E.A.D., and P.A.L. conceived the project. M.H.M. constructed the theoretical framework with D.D.S., E.A.D., and P.A.L.. D.D.S. set-up and carried out the numerical simulations. All authors contributed in analyzing and interpreting the data. All authors contributed in writing the paper.

\section{Competing interests}

The authors declare no competing interests.

\onecolumngrid

\setcounter{figure}{0}
\setcounter{equation}{0}
\renewcommand{\theequation}{A.\arabic{equation}}
\renewcommand\thefigure{A.\arabic{figure}}

\section{Appendix 1: Electrodynamics of the pseudogap bilayer $\rm YBCO$}
\label{Model}

\begin{wrapfigure}[24]{L}{0.25\textwidth}
\centering
\vspace{-0.3cm}
\includegraphics[trim = 5 10 5 18, clip, scale = 1 ]{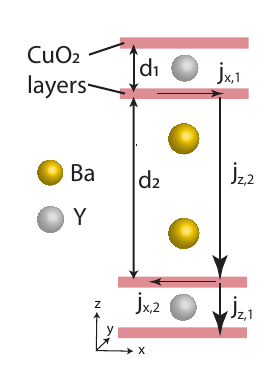}
  \caption{\textbf{Sketch of YBCO.} The crystal structure of YBCO is made up of pairs of $\rm CuO_2$-layers that below $T_c$ become superconducting. Members of a $\rm CuO_2$ bilayer are separated by an Yttrium atom and an interlayer distance of $d_1 = \SI{4}{\angstrom}$, while different bilayers are separated by two Barrium atoms and a distance of $d_2 = \SI{8}{\angstrom}$. In the figure we also annotated the position of the in-plane and out-of plane currents. } 
  \label{fig:YBCOsketch}
\end{wrapfigure}

\paragraph{Maxwell equations for the layered YBCO.} As sketched in Fig.~\ref{fig:YBCOsketch}, the YBCO structure consists of bilayers with spacing $d_1$, which are separated from each other by a larger spacing $d_2$. The unit cell dimension along the $z$-direction is $d_1 + d_2$. We shall employ the layer index $i = \{1, 2\}$ to denote the bottom and top members of a bilayer, and $l$ to denote the $l$-th unit cell. We start our description by considering the Maxwell equations:
\begin{align}
   \frac{1}{c^2}\partial_t E_z + \frac{1}{c^2 \epsilon_0} j_z =& \, \partial_x B_y, \label{eq:Max1} \\
   \frac{1}{c^2}\partial_t E_x + \frac{1}{c^2\epsilon_0} j_x =& \, \partial_z B_y, \label{eq:Max2}\\
   B_y =& \, \partial_z A_x - \partial_x A_z ,
\end{align}
where we take the system to be homogeneous along the $y$-axis. To model the low-energy electrodynamics of the YBCO system, we assume that physical quantities are approximately constant between $\rm{CuO}$ layers and can be replaced by their averages. For the $z$-component of the electric field, we thus focus on:
\begin{align}
    E_{z,1,l} =& \, \frac{1}{d_1}\int^{ z_{2,l}}_{z_{1,l} } dz E_z (z), \\
    E_{z,2,l} =& \, \frac{1}{d_2} \int_{ z_{2,l}}^{z_{1,l+1}} dz E_z(z) .
\end{align}
Similarly, for the $y$-component of the magnetic field we have that:
\begin{align}
    B_{y,1,l} =& \, \frac{A_{x,2,l} - A_{x,1,l}}{d_1} - \partial_x A_{z,1,l},\label{eq:MagnAx1}\\
    B_{y,2,l} =& \, \frac{A_{x,1,l} - A_{x,2,l}}{d_2} - \partial_x A_{z,2,l} \label{eq:MagnAx2},
\end{align}
where we have defined $A_{x,1,l} = A_{x}(z = z_{1,l}) $, $A_{x,2,l} = A_{x}(z = z_{2,l})$, $A_{z,1,l} = \frac{1}{d_1} \int_{ z_{1,l}}^{z_{2,l}} dz A_z(z) $, and $A_{z,2,l} = \frac{1}{d_2} \int_{ z_{2,l}}^{z_{1,l+1}} dz A_z(z)$. Moreover, we assume that the current along the $x$-direction is localised at each $\rm{CuO}$ layer: 
\begin{equation}
    j_x(z) = \sum_l \left[ j_{x,1,l} \delta(z - z_{1,l}) + j_{x,2,l} \delta (z - z_{2,l}) \right].
\end{equation}

Integrating Eq.~\eqref{eq:Max1} over $d_1$ (distance between layers) and Eq.~\eqref{eq:Max2} across the layer thickness, we find a set of closed equations in terms of the intra-bilayer and inter-bilayer fields:
\begin{align}
   \frac{1}{c^2}\partial_t E_{z,1,l} + \frac{1}{c^2 \epsilon_0} j_{z,1,l} =& \, \partial_x B_{y,1,l}, \label{eq:MaxzAver1}\\
   \frac{1}{c^2}\partial_t E_{z,2,l} + \frac{1}{c^2 \epsilon_0} j_{z,2,l} =& \, \partial_x B_{y,2,l}, \label{eq:MaxzAver2} \\
   \frac{1}{c^2 \epsilon_0} j_{x,1,l} =& \, B_{y,2,l-1} - B_{y,1,l}  , \label{eq:Maxcurr1}\\
   \frac{1}{c^2 \epsilon_0} j_{x,2,l} =& \, B_{y,1,l} - B_{y,2,l} , \label{eq:Maxcurr2}
\end{align}
where the average currents along the $z$-direction are given by $j_{z,1,l} = \frac{1}{d_1} \int_{z_{1,l}}^{z_{2,l}} dz \, j_z(z)$ and $j_{z,2,l} = \frac{1}{d_2} \int_{z_{2,l}}^{z_{1,l+1}} dz \, j_z(z)$.

\paragraph{Dissipationless counterflow model for pseudogap YBCO.} The dissipationless counterflow is modelled by a superfluid response in plane, with a friction force created by the motion of pancake vortices\cite{Nelson79}. According to Ref.~\citenum{Homann24}, vortices flow only due to the sum of the currents in the two bilayers. When the currents are anti-symmetric, intra-bilayer coherence can pin the vortices in the two layers, allowing the counterflow current to remain dissipationless. This scenario is captured by the superfluid equations:
\begin{align}
    \partial_t j^s_{z,1,l} =& \, \frac{e^*}{\hbar} d_1 J_{c,1} \cos(\theta_{1,l}) E_{z,1,l}, \\
    \partial_t j^s_{z,2,l} =& \, \frac{e^*}{\hbar} d_2 J_{c,2} \cos(\theta_{2,l}) E_{z,2,l} , \\
    \partial_t j^s_{x,1,l} =& \, \frac{(e^*)^2 n_s}{m} E_{x,1,l} + \gamma_d \left( j^s_{x,1,l} + j^s_{x,2,l} \right), \label{eq:CurrxSuper1}\\
    \partial_t j^s_{x,2,l} =& \, \frac{(e^*)^2 n_s}{m} E_{x,2,l} + \gamma_d \left( j^s_{x,1,l} + j^s_{x,2,l} \right), \label{eq:CurrxSuper2}
\end{align}
where $\theta_{1,l} = \phi_{2,l} - \phi_{1,l} - \frac{e^*}{\hbar} d_1 A_{z,1,l}$ and $\theta_{2,l} = \phi_{1,l+1} - \phi_{2,l} - \frac{e^*}{\hbar} d_2 A_{z,2,l}$ are the gauge invariant phases, directly related to the electric fields through the relations $\hbar \partial_t \theta_{1,l} = e^* d_1 E_{z,1,l}$ and $\hbar \partial_t \theta_{2,l} = e^* d_1 E_{z,2,l}$, $E_{x,1,l} = E_x(z = z_{1,l})$ and $E_{x,2,l} = E_x(z = z_{2,l})$ are the in-plane electric fields, $n_s$ is the 2D superfluid density in each $\rm{CuO}$-layer, and $\gamma_d$ is a friction coefficient accounting for the vortex motion in the presence of a symmetric current in each bilayer. In the counterflow state, we take $J_{c,2} = 0$ and assume that the bilayer Josephson coupling is unaffected, i.e., $J_{c,1} = J_{c,1}(T= 0 \, \rm{K})$.

We henceforth examine the system's response at momentum $q_z = 0$, where the fields have the same amplitude in each unit cell, and drop the $l$-index dependence. Equations~\eqref{eq:Maxcurr1} and \eqref{eq:Maxcurr2} always imply equal and opposite in-plane currents, $j_{x,2} = - j_{x,1}$, thereby leading to dissipationless in-plane relative phase dynamics. 
Dissipation is nevertheless included as $j^n_{z,1} = \sigma_{n,1} E_{z,1}$ and $j^n_{z,2} = \sigma_{n,2} E_{z,2}$, where $\sigma_{n,1}$ is the normal c-axis bilayer conductivity and $\sigma_{n,2}$ is a similar quantity motivated by the cavity to the radiation field formed by the region in-between bilayers being leaky. The cavity Q leads to dissipation of photon modes, which can be modeled by adding a term proportional to the intra-bilayer electric field in Eq.~\eqref{eq:MaxzAver2}. This will lead to a dissipative term parametrized by $\sigma_{n,2}$ in the equation for $\theta_2$ below. Note that $\sigma_{n,2}$ is a parameter that models the cavity Q and should not be interpreted as conductivity between bilayers.

Equations~\eqref{eq:Maxcurr1}--\eqref{eq:Maxcurr2} express the discontinuity of the magnetic field $B_y$ across each layer in terms of the in-plane current. The latter can be formulated through $\theta$ and $B_y$ by writing the electric field along the $x$-direction in terms of the gauge invariant phase, $\partial_t \left( \hbar \partial_x \phi_1 - e^* A_{x,1}\right) = e^* E_{x,1} $ and $\partial_t \left( \hbar \partial_x \phi_2 - e^* A_{x,2}\right) = e^* E_{x,2}$, and using Eqs.~\eqref{eq:CurrxSuper1}--\eqref{eq:CurrxSuper2} and Eqs.~\eqref{eq:MagnAx1}--\eqref{eq:MagnAx2}. Equations~\eqref{eq:Maxcurr1}--\eqref{eq:Maxcurr2} become
\begin{align}
  \beta \left(  \partial_x \theta_1 - \frac{e^*}{\hbar} d_1 B_{y,1} \right) = & \, \frac{e^*}{\hbar} d_1 \left( B_{y,1} - B_{y,2} \right) ,\label{eq:MagnTheta1}\\
  \frac{d_2}{d_1} \beta \left( \partial_x \theta_2 -\frac{e^*}{\hbar} d_2 B_{y,2} \right) = & \, \frac{e^*}{\hbar} d_2 \left( B_{y,2} - B_{y,1} \right)  
 \label{eq:MagnTheta2}
\end{align}
where $\beta = \frac{\left(e^*\right)^2 n_s d_1 }{2 \epsilon_0 m c^2}$. We use Eqs.~\eqref{eq:MagnTheta1} and \eqref{eq:MagnTheta2} to solve for $B_{y,1}$ and $B_{y,2}$ in terms of $\theta_1$ and $\theta_2$. Plugging into Eqs.~\eqref{eq:MaxzAver1} and \eqref{eq:MaxzAver2}, we arrive at the equations of motion for the gauge invariant phases in the pseudogap counterflow state:
%\vspace{4mm}
\begin{align}
    \frac{1}{c^2} \partial_t^2 \theta_1+ \frac{\omega^2_{J,1}}{c^2} \sin \theta_1 + \frac{\sigma_{n,1}}{\epsilon_0 c^2} \partial_t \theta_1 - \frac{1}{d_2(1 + \beta) + d_1 } \partial_x^2 \left[ (d_1+ d_2 \beta ) \theta_1 + d_1 \theta_2\right]  = & \, 0, \label{eq:Layer1EOM} \\
    \frac{1}{c^2} \partial_t^2 \theta_2+ \frac{\sigma_{n,2}}{\epsilon_0 c^2} \partial_t \theta_2- \frac{1}{d_2(1 + \beta) + d_1} \partial_x^2 \left[ d_2 (1 + \beta ) \theta_2 + d_2 \theta_1 \right] = & \, 0,
\label{eq:Layer2EOM}
\end{align}
which depend on the frequency of the upper plasmon $\omega_{J,1} \approx 2 \pi \times 14 $ THz, the intra-bilayer distance $d_1 \approx$  \SI{4}{\angstrom}, the inter-bilayer distance $d_2 \approx$ \SI{8}{\angstrom}, and the dimensionless parameter $\beta$. In Appendix~2, we show that the value $\beta \approx 10^{-5}$ can be estimated from experiments as the diamagnetic response of the pseudogap phase due to counterflow screening.

\paragraph{External driving.} In the presence of an homogeneous pump electric field along the $z$-direction, $E_{p} = E_0 f(t)$, the currents along the $z$-direction are given by: 
\begin{align}
    j^s_{z,1} =& \, J_{c,1} \sin\left(\theta_1 - \frac{e^*}{\hbar} d_1 A_{p,1} \right) , \\
    j^n_{z,i} =& \, \frac{\sigma_{n,i}}{e^* d_i} \partial_t \left(\theta_i - \frac{e^*}{\hbar} d_i A_{p,i}\right), 
\end{align}
where $i = \{1,2\}$, $ \partial_t A_{p,1} = - E_{p} $, and $ \partial_t A_{p,2} = - E_{p} $. This leads to the driven equations of motion: 
\begin{align}
    \frac{1}{c^2} \partial_t^2 \theta_1+ \frac{\omega^2_{J,1}}{c^2} \sin\left(\theta_1 - \frac{e^*}{\hbar} d_1 A_{p,1}\right) + \frac{\sigma_{n,1}}{\epsilon_0 c^2} \partial_t\left(\theta_1 - \frac{e^*}{\hbar} d_1 A_{p,1}\right) - \frac{1}{d_2(1 + \beta) + d_1 } \partial_x^2 \left[ (d_1+ d_2 \beta ) \theta_1 + d_1 \theta_2\right]  =& \, 0, \\
    \frac{1}{c^2} \partial_t^2 \theta_2+ \frac{\sigma_{n,2}}{\epsilon_0 c^2} \partial_t\left(\theta_2 - \frac{e^*}{\hbar} d_2 A_{p,2}\right)- \frac{1}{d_2(1 + \beta) + d_1} \partial_x^2 \left[ d_2 (1 + \beta ) \theta_2 + d_2 \theta_1 \right] =& \, 0.
\end{align}
The latter equations can be expressed in a simpler form via the uniform transformation $\theta_1 = \Theta_1 + \frac{e^*}{\hbar} d_1 A_{p,1}  $ and $\theta_2 = \Theta_2 + \frac{e^*}{\hbar} d_2 A_{p,2}$ : 
\begin{align}
    \frac{1}{c^2} \partial_t^2 \Theta_1+ \frac{\omega^2_{J,1}}{c^2} \sin \Theta_1 + \frac{\sigma_{n,1}}{\epsilon_0 c^2} \partial_t \Theta_1 - \frac{1}{d_2(1 + \beta) + d_1 } \partial_x^2 \left[ (d_1+ d_2 \beta ) \Theta_1 + d_1 \Theta_2\right]  =& \, \frac{e^* d_1}{\hbar c^2} \partial_t E_p(t), \label{app:eq:DrivenMulti1}\\
    \frac{1}{c^2} \partial_t^2 \Theta_2+ \frac{\sigma_{n,2}}{\epsilon_0 c^2} \partial_t \Theta_2 - \frac{1}{d_2(1 + \beta) + d_1} \partial_x^2 \left[ d_2 (1 + \beta ) \Theta_2 + d_2 \Theta_1 \right] =& \, \frac{e^* d_2}{\hbar c^2} \partial_t E_p(t).
\label{app:eq:DrivenMulti2}
\end{align}
In this basis, the equations of motion resemble that solved for coupled systems of driven damped long Josephson junctions\cite{Likharev86,Ustinov98,Cuevas14,DeSantis23,DeSantis24,DeSantis25}.

\paragraph{Boundary conditions for a periodic bilayer system of finite width.} To account for the physics between two cracks, it is necessary to define the boundary conditions that simulate the electrodynamics outside the pumped finite segment. Indeed, as local paramagnetic currents are formed and amplified within the pre-soliton regime, an equal and opposite flux is anticipated to stem from the boundaries, ensuring conservation of the total flux. The boundary conditions at the $x = 0$ interface are expressed as follows\cite{Marios22}:
\begin{align}
    \left. E_{z,\mathrm{air}, q_z = 0}\right|_{x = 0} =& \left. E_{z,\mathrm{mat}, q_z = 0}\right|_{x = 0}, \\
    \left. B_{y,\mathrm{air}, q_z = 0}\right|_{x = 0} =& \left. B_{y,\mathrm{mat}, q_z = 0}\right|_{x = 0}, \\
    \left. j_{x,1}\right|_{x = 0} \propto & \left. \left( B_{y,2} - B_{y,1} \right) \right|_{x = 0} = 0,
\end{align}
where the first two conditions state that the tangential components of the average electric and magnetic fields are continuous across the interface. We break up the fields in air as externally imposed contributions plus propagating modes: 
\begin{align}
    \left. E_{z,\mathrm{air},q_z = 0}\right|_{x = 0} =& \left. E_{\mathrm{air},q_x}\right|_{x = 0} + E_{p},\\
    \left. B_{y,\mathrm{air},q_z = 0}\right|_{x = 0} =& \left. B_{\mathrm{air}, q_x}\right|_{x = 0} + B_{\rm{ext}}.
\end{align}
For a left (right) propagating mode, we have the condition $B_{\mathrm{air},q_x} = + \, (-) \, \frac{1}{c} E_{\mathrm{air},q_x}$, implying: 
\begin{equation}
    \left. B_{y,\mathrm{mat},q_z = 0}\right|_{x = 0} - B_{\rm{ext}}  = + \, (-) \, \frac{1}{c} \left[ \left. E_{z, \mathrm{mat}, q_z = 0}\right|_{x = 0} - E_{p} \right] .
\end{equation}
Taking the average electric field as $E_{z,\mathrm{mat},q_z = 0} = \hbar \frac{\partial_t \Theta_1 + \partial_t \Theta_2}{e^* ( d_1 + d_2 )} = \hbar \frac{\partial_t \theta_1 + \partial_t \theta_2}{e^* ( d_1 + d_2 ) } + E_{p}$, we get the following boundary conditions:
\begin{align}
    \left. \left\lbrace \partial_x \Theta_1 + \partial_x \Theta_2 \mp \frac{1}{c} \left[ \partial_t \Theta_1 + \partial_t \Theta_2 - e^* (d_1 + d_2)E_{p} \right] \right\rbrace \right|_{x=0,L} =& \, \frac{e^* (d_1 + d_2)}{\hbar} B_{\rm{ext}}, \label{eq:BCMulti1} \\ 
    \left. \left( \frac{\partial_x \Theta_1}{d_1} - \frac{\partial_x \Theta_2}{d_2} \right) \right|_{x=0,L} =& \, 0, \label{eq:BCMulti2}
\end{align}
where the $-$ ($+$) sign applies to the $x = 0$ ($x = L$) boundary. Throughout this work, we report Eqs.~\eqref{app:eq:DrivenMulti1}--\eqref{app:eq:DrivenMulti2} and Eqs.~\eqref{eq:BCMulti1}--\eqref{eq:BCMulti2} and drop the capitals, i.e., we redefine $\Theta_1 \equiv \theta_1$ and $\Theta_2 \equiv \theta_2$.

\paragraph{Single bilayer limit.} Above, we have derived the electromagnetic equations of motion for a multi-bilayer system. A physically instructive limit to take is that of a single bilayer. To this end, we let $d_2 \rightarrow \infty$, obtaining the intra-bilayer equation:
\begin{equation}
     \frac{1}{c^2} \partial_t^2 \theta+ \frac{\omega^2_{J}}{c^2} \sin \theta + \frac{\sigma_{n}}{\epsilon_0 c^2} \partial_t \theta - \frac{\beta}{1+ \beta} \partial_x^2  \theta  = \frac{e^* d_1}{\hbar c^2}\partial_t E_p ,
\end{equation}
where (as in the main text) we suppressed the `1' subscript: $\theta_1 \equiv \theta$, $\omega_{J,1} \equiv \omega_J$, and $\sigma_{n,1} \equiv \sigma_n$. As the previous equation indicates, in this limit, the bilayer dynamics is decoupled from that in-between the bilayers. The resulting equation, referred to in the literature as the perturbed (or driven damped) SG equation\cite{Cuevas14,DeSantis23,DeSantis24,DeSantis25}, captures the instabilities discussed in the main text. The multi-bilayer solution then yields corrections to the SG prediction, accounting for the capacitive coupling between layers. The inter-bilayer dynamics is described by:
\begin{equation}
     \partial_t^2 \theta_2 + \frac{\sigma_{n,2}}{\epsilon_0} \partial_t \theta_2 - c^2 \partial_x^2 \theta_2 = \frac{c^2}{1 + \beta} \partial_x^2 \theta +  \frac{e^* d_2}{\hbar} \partial_t E_p. 
\end{equation}
This highlights that capacitive coupling between neighboring layers remains significant in the present limit and reinforces the observation that, for large but finite $d_2$, the current generated within the bilayer is converted into a displacement current between bilayers, resulting in the overall edge current discussed in the main text.

\section{Appendix 2: Equilibrium counterflow diamagnetism}
\label{app:equilCounter}

Here we consider a semi-infinite geometry with a interface to air at $x = 0$. In the presence of a magnetic field, the boundary conditions are given by $ \left. B_{y,1} \right|_{x = 0}  = \left. B_{y,2}\right|_{x = 0} = B_{\rm{ext}}$. In terms of the gauge invariant phases, we get:
\begin{align}
    \left. \partial_x \theta_1 \right|_{x = 0} =& \, \frac{e^* d_1}{\hbar} B_{\rm{ext}} ,\\
    \left. \partial_x \theta_2 \right|_{x = 0} =& \, \frac{e^* d_2}{\hbar} B_{\rm{ext}} .
\end{align}
To analytically solve the equilibrium problem, it is simpler work directly with the magnetic fields, which using Eqs.~\eqref{eq:MagnTheta1} and \eqref{eq:MagnTheta2} are found to satisfy:
\begin{align}
    \partial_x^2 B_{y,1} =& \left( \frac{\omega^2_{J,1}}{c^2} \cos \theta_1 + \frac{1}{c^2}\partial^2_t + \frac{\sigma_{n,1}}{\epsilon_0 c^2} \partial_t\right)\left(  \frac{B_{y,1} - B_{y,2}}{\beta} + B_{y,1}  \right)  ,\label{eq:EoMB1} \\
    \partial^2_x B_{y,2} =& \left( \frac{1}{c^2}\partial^2_t + \frac{\sigma_{n,2}}{c^2} \partial_t \right)\left( \frac{B_{y,2} - B_{y,1}}{\beta\frac{d_2}{d_1} } + B_{y,2} \right) .\label{eq:EoMB2}
\end{align}
In the static limit $\theta_1$ is small, such that $\cos \theta_1 \approx 1$, and our equations become:
\begin{align}
    \partial_x^2 B_{y,1} =& \, \frac{\omega^2_{J,1}}{c^2} \left(  \frac{B_{y,1} - B_{y,2}}{\beta} + B_{y,1}  \right), \label{eq:EoMB1static} \\
    \partial^2_x B_{y,2} =& \, 0. \label{eq:EoMB2static}
\end{align}
Solutions of the latest system of equations are readily obtained as:
\begin{align}
    B_{y,1}(x) =& \, \left( B_{\rm{ext}} - B_{\rm{sc}} \right) e^{- k x} + B_{\rm{sc}}, \\
    B_{y,2}(x) =& \, B_{\rm{ext}},
\end{align}
where $B_{\rm{sc}} = \frac{1}{1 + \beta} B_{\rm{ext}} $ is the screened field and $ k = \lambda^{-1} = \frac{\omega_{J,1}}{c}\sqrt{\frac{1 + \beta}{\beta}}$ is the inverse penetration depth. These solutions imply that, inside the bilayer, we have a diamagnetic susceptibility given by $ \chi_{\rm{bilayer}} = - \beta $. Since the bilayer makes up $1/3$ of the unit cell, the effective diamagnetic susceptibility of the entire structure due to counterflow currents is given by:
\begin{equation}
   \chi_{d} = - \frac{\beta}{3} .
\end{equation}
According to Ref.~\citenum{Cooper12}, which reports on measurements of magnetic properties of YBCO, there exists an excess diamagnetic contribution of $ \chi_{d} = - 10^{-5}$ in the pseudogap regime. Assuming that such value originates entirely from the counterflow diamagnetism described here, we arrive to $\beta = 3 \times 10^{-5}$.

In the single bilayer limit, $d_2 \rightarrow \infty $, the following expression holds for the magnetic field $B_{y,1}$:
\begin{equation}
    B_{y,2} - B_{y,1}(x) = \frac{\beta}{1 + \beta} \left( B_{\rm{ext}} - \frac{\hbar}{e^* d_1} \partial_x \theta_1 \right) 
\end{equation}
and the in-plane screening currents are:
\begin{equation}
     \frac{1}{c^2 \epsilon_0} j_{x,1} = - \frac{1}{c^2 \epsilon_0} j_{x,2} = B_{y,2} - B_{y,1} = \frac{\beta}{1 + \beta} \left( B_{\rm{ext}} - \frac{\hbar}{e^* d_1} \partial_x \theta_1 \right) .
\end{equation}

\section{Appendix 3: Floquet effective theory for the SG pre-soliton regime}
\label{SGFloquet}

We decompose the solutions of the perturbed SG model as  $\theta(x,t) = \tilde{\theta}(t) + \theta'(x,t)$, where $\tilde{\theta}$ is a spatially homogeneous oscillating component obeying the driven damped pendulum equation (in rescaled units, $x \to x / \lambda$ and $t \to t \, \omega_{J}$, see the main text):
\begin{equation}
    \partial_t^2 \tilde{\theta} + \alpha \partial_t \tilde{\theta} + \sin \tilde{\theta} = \partial_t V(t),
\end{equation}
while $\theta'$ satisfies the inhomogeneous boundary conditions and is subject to the effective SG potential created by the oscillations of $\tilde{\theta}$: 
\begin{align}
    \partial_t^2 \theta' + \alpha \partial_t \theta' + \cos \tilde{\theta} \sin \theta' + (\cos \theta' - 1) \sin \tilde{\theta} =& \, 0 ,\\
    \left. \partial_x \theta'\right|_{x=0,L} =& \, q.
\end{align}
We then approximate $\tilde{\theta}(t)$ as a function oscillating with frequency $\omega_p$, but with a slowly time-varying amplitude (that is, on the timescale of the pump's envelope):
\begin{equation}
    \tilde{\theta}(t) = A(t) \sin(\omega_p t) . 
\end{equation}
For a fixed amplitude $A$, we can expand the nonlinear SG potential into harmonics oscillating at multiples of $\omega_p$:
\begin{equation}
    \cos \left[ A \sin(\omega_p t) \right] = J_0(A) + \sum_n J_{2n}(A) e^{- i 2 n \omega_p t}.
\end{equation}
Even if the pump's amplitude is very large, in the range of small $q$ of interest for experimental purposes (see the main text), the dynamical equation for $\theta'$ can be linearized. The resulting effective equation of motion, in frequency space, takes the form:
\begin{equation}
     \left[ - \omega^2 + i \alpha \omega + J_0(A) \right] \theta'(\omega) + \sum_n \left[ J_{2n}(A) \theta'(\omega + 2n \omega_p) + J_{-2n}\theta'(\omega - 2 n \omega_p) \right] = 0 ,
\end{equation}
where different frequency components of $\theta'$ are coupled to all the Floquet copies. Restricting $\omega$ to the first Brillouin zone of the Floquet problem, $\omega \in (-\omega_p /2 , \omega_p /2)$, we can express different Floquet components using the notation, $\theta'^{(n)} = \theta'(\omega + n \omega_p)$. By means of a Floquet perturbation theory\cite{Eckardt_2015}, we can include the effects of higher $\theta'$ harmonics on the dynamics of the slowly varying $\theta'^{(0)}(\omega)$, up to second order in the off-diagonal components $J_{2n}(A)$. Assuming $\omega \ll \omega_p$, the equations of motion become, to leading order:
\begin{align}
    &\left[ - \omega^2 + i \alpha \omega
 + J_0(A)\right]\theta'^{(0)} + J_{2n}(A) \theta'^{(2n)} + J_{-2n}(A)\theta'^{(-2n)} = 0 , \\
&\left[ - (2 n \omega_p)^2 + i \alpha 2 n \omega_p
 + J_0(A)\right] \theta'^{(2n)} = - J_{-2n}(A) \theta'^{(0)}, 
 \end{align}
Combining the two equations together, we find the effective equation for $\theta'^{(0)}$, to second order in the Floquet expansion:
\begin{align}
\begin{split}
 &\left[ - \omega^2 + i \alpha \omega
 + J_0(A)\right] \theta'^{(0)} - J_{2n}(A) J_{-2n}(A) \left[ \frac{1}{ - (2 n \omega_p)^2 + i \alpha 2 n \omega_p
 + J_0(A)} + \frac{1}{ - (2 n \omega_p)^2 - i \alpha 2 n \omega_p
 + J_0(A)} \right] \theta'^{(0)} = 0 ,
 \end{split}\\
\begin{split}
 &\left[ - \omega^2 + i \alpha \omega
 + J_0(A)\right] \theta'^{(0)} + 2 J_{2n}(A) J_{-2n}(A) \left\lbrace \frac{(2n \omega_p)^2 - J_0(A)}{\left[ (2n \omega_p)^2 - J_0(A) \right]^2 + \alpha^2 (2 n \omega_p)^2 }  \right\rbrace \theta'^{(0)} = 0 .
 \end{split}
\end{align}
This leads to a parametric drive term $\eta$:
\begin{equation}
    \eta(A) = J_0(A) + 2 \sum_n J_{2n}(A) J_{-2n}(A) \left\lbrace \frac{(2n \omega_p)^2 - J_0(A)}{\left[ (2n \omega_p)^2 - J_0(A)\right]^2 + \alpha^2 (2 n \omega_p)^2 } \right\rbrace .
\end{equation}
In practice, we solve for $\tilde{\theta}$ numerically, and then extract its envelope via a Hilbert transform. Due to the nonlinear nature of the pendulum equation, $\tilde{\theta}$ is not described just by a single harmonic, $\tilde{\theta} = A \sin(\omega_p t) $, but in general higher harmonic contributions will come into play. Indeed, the envelope we compute usually presents some fast components on top of a smooth background. We neglect such higher harmonic contributions by employing a low-pass filter. We find this scheme to provide a simple yet accurate approximation for most of our purposes; however, at very large pump amplitudes, the neglected higher harmonics should also be included in the Floquet analysis.

\section{Appendix 4: Dynamical SG response}
\label{SG}

\begin{figure*}[t!]
    \centering
    \includegraphics[scale = 1]{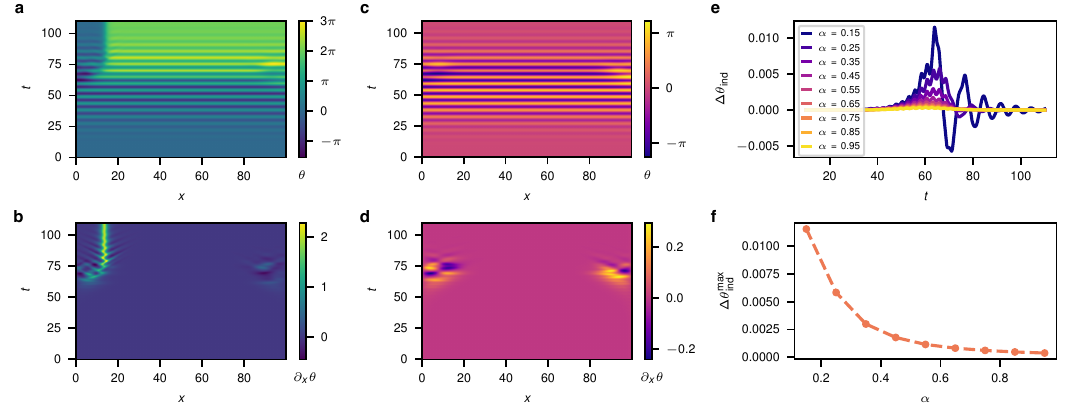}
    \caption{\textbf{Spatio-temporal behaviors and damping dependence in the driven SG model.} \textbf{a} Contour plot of the phase and \textbf{b} contour plot of the phase gradient for the (normalized) parameter combination $V_0 = 4$ and $q = 0.01$, representative of the soliton regime, discussed in the main text. Under driving, small boundary inhomogeneities caused by nonzero-flux boundary conditions gradually increase until a soliton, located around $x = 10$ in this example, is excited. \textbf{c} Contour plot of the phase and \textbf{d} contour plot of the phase gradient for the combination $V_0 = 3.5$ and $q = 0.01$, taken in the pre-soliton scenario. An edge-localized excitation, qualitatively similar to a soliton but with much smaller amplitude, develops during pumping. The following (dimensionless) parameters are used in panels (a)-(d): $L=100$, $\alpha=0.5$, and $\beta = 10^{-5}$. \textbf{e} Time traces of the photo-induced phase difference across the entire segment and \textbf{f} maximum photo-induced phase difference versus the damping coefficient, for $V_0 = 2.5$, $q = 1.4 \times 10^{-4}$, $L=100$, and $\beta = 10^{-5}$. As dissipation decreases, we observe an enhancement in the system's peak pre-solitonic response. The lower dissipation scenario, however, is also characterized by more pronounced oscillatory transients.}
    \label{fig:SuppInfoSG}
\end{figure*}
\paragraph{Dynamically excited SG solitons and pre-solitons.} In Fig.~\ref{fig:SuppInfoSG}(a) and (b), we present contour plots of the phase and its gradient, respectively, to illustrate how solitons can enter the system through the cooperative action of a small boundary field and strong spatially uniform driving. The latter induces an fast oscillating background, over which initial boundary-localized profiles surf and get amplified. This eventually leads to the penetration of a full soliton, located at $x \approx 10$ in the shown example, as indicated by the $2 \pi$ phase step or equivalently by the peak gradient response of roughly $2$. We note in passing that the soliton, if formed sufficiently away from the edge, persists long after the pump pulse has passed by virtue of its topological robustness. It is also worth mentioning that the strongly nonlinear (soliton) regime---in particular, the solitonic content of the final state after pumping---depends on the phase of the oscillating pump field. A $\pi$-shifted pump oscillation would result in a soliton emerging from the opposite edge, effectively reversing the outcome reported here.

In Fig.~\ref{fig:SuppInfoSG}(c) and (d), we show the spatio-temporal phase and gradient profiles for a case representative of the pre-soliton regime. Here, in the presence of strong driving, a soliton-like structure emerges out of the equilibrium configuration, which is characterized by very small gradients at the edges and is zero elsewhere, surfing over the (pump-induced) uniform oscillations. In this scenario, the localized feature we refer to as pre-soliton has a much smaller amplitude than an actual SG soliton, and it fades away once the pump pulse is gone at a rate influenced by the damping coefficient.

\paragraph{Damping dependence within the SG pre-soliton regime.} We now address the sensitivity of our SG pre-soliton results to damping, by varying the $\alpha$ parameter over an order of magnitude: $\alpha \in (0.1, 1)$. As shown in Fig.~\ref{fig:SuppInfoSG}(e), the time traces of the photo-induced phase difference indicate a more pronounced overall response as dissipation is decreased, with both larger peak values and oscillations on the way back to equilibrium. It should be pointed out that, to precisely illustrate the dynamical traits under discussion, here we do not perform any convolution operation on the simulation results, as opposed to what is mostly done elsewhere in this work (e.g., in the main text's plots, oscillations at a period of $\sim 1 / \omega_p$ during the pre-solitonic build-up are filtered out). The results' dependence on $\alpha$ is further highlighted by considering the peak response, see Fig.~\ref{fig:SuppInfoSG}(f). In this case, we observe a power-law-type attenuation of the maximum photo-induced phase difference with the dissipation strength. In general, as previously hinted in our work, the $\alpha \to 0$ limit can be rather subtle: as exploring the full SG excitation spectrum progressively becomes easier, complications to the simple instability picture painted above can gradually arise. For example, the emergence of long-lived (for $\alpha \to 0$) breather-like modes can enhance the oscillatory character of the $\Delta \theta_{\rm{ind}}$ time traces, as compared to the mostly positive result obtained in the medium-to-large dissipation range.

\section{Appendix 5: Floquet multi-bilayer response}
\label{Multi}

\paragraph{Multi-bilayer Floquet effective framework.} We start from Eqs.~\eqref{app:eq:DrivenMulti1} and \eqref{app:eq:DrivenMulti2} derived above, which we now write in rescaled units, $x \to x / \lambda$ and $t \to t \, \omega_{J,1}$:
\begin{align}
    \partial_t^2 \theta_1+ \sin \theta_1 + \alpha_1 \partial_t \theta_1 - \frac{1 + \beta}{\beta} \partial_x^2 \left( K_1^+ \theta_1 + K_2 \theta_2 \right)  =& \, \partial_t V(t) , \label{eq:DrivenMulti1Scaled}\\
    \partial_t^2 \theta_2+ \alpha_2 \partial_t \theta_2 - \frac{1 + \beta}{\beta} \partial_x^2 \left( K_2^+ \theta_2 + K_1 \theta_1 \right) =& \, \frac{d_2}{d_1} \partial_t V(t) ,
\label{eq:DrivenMulti2Scaled}
\end{align}
where $K_1^+ = \frac{d_1+ d_2 \beta }{d_2(1 + \beta) + d_1 }$, $K_2 = \frac{d_1 }{d_2(1 + \beta) + d_1 }$, $K_2^+ = \frac{d_2 ( 1 + \beta ) }{d_2(1 + \beta) + d_1 }$, and $K_1 = \frac{d_2 }{d_2(1 + \beta) + d_1 }$. We decompose our phase profiles as $\theta_{1,2}(x,t) = \tilde{\theta}_{1,2}(t) + \theta'_{1,2}(x,t)$, with the spatially homogeneous components defined as the solutions of the following equations:
\begin{align}
    \partial_t^2 \tilde{\theta}_1 + \sin \tilde{\theta}_1 + \alpha_1 \partial_t \tilde{\theta}_1 &= \, \partial_t V(t), \\
    \partial_t^2 \tilde{\theta}_2 + \alpha_2 \partial_t \tilde{\theta}_2 &= \, \frac{d_2}{d_1} \partial_t V(t),
\end{align}
and the boundary-condition-sensitive components obeying:
\begin{align}
    \partial_t^2 \theta'_1+ \eta \theta'_1 + \alpha_1 \partial_t \theta'_1 - \frac{1 + \beta}{\beta} \partial_x^2 \left( K_1^+ \theta'_1 + K_2 \theta'_2 \right)  =& \, 0 , \label{eq:Multi1Floquet}\\
    \partial_t^2 \theta'_2+ \alpha_2 \partial_t \theta'_2 - \frac{1 + \beta}{\beta} \partial_x^2 \left( K_2^+ \theta'_2 + K_1 \theta'_1 \right) =& \, 0 ,
\label{eq:Multi2Floquet}
\end{align}
where the equation of motion of $\theta'_1$ is linearized via the Floquet approach illustrated above, such that $\eta$ is a parametric drive term that we extract from the driven damped pendulum solution. We evaluate the ratio between the $z$-averaged magnetic field within the unit cell and the equilibrium diamagnetic field amplitude as:
\begin{equation}
    \frac{B_y}{\left| B_{\rm{ext}} \chi_d \right|} = \frac{d_1 B_{y,1} + d_2 B_{y,2}}{\left( d_1 + d_2 \right) \left| B_{\rm{ext}} \chi_d \right|} = \frac{d_1}{d_1 + d_2} \frac{\partial_x \theta'_1 + \partial_x \theta'_2}{\left| q \chi_d \right|} .
\end{equation}
We can conveniently apply a shift to the inter-bilayer phase gradient, $ \partial_x \theta'_2 \to \partial_x \theta'_2 - q \left( \frac{d_2}{d_1} + \frac{1}{1 + \beta} \right) $, which does not influence the dynamical equations, so that we get:
\begin{equation}
    \frac{d_1}{d_1 + d_2} \frac{\partial_x \theta'_1 + \partial_x \theta'_2}{\left| q \chi_d \right|} = \frac{d_1 \left( B_{y,1} - B_{\rm{sc}} \right) + d_2 \left( B_{y,2} - B_{\rm{ext}} \right)}{\left( d_1 + d_2 \right) \left| B_{\rm{ext}} \chi_d \right|} \sim \frac{B_{\rm{ind}}}{\left| B_{\rm{ext}} \chi_d \right|} . \label{eq:ratioB}
\end{equation}
In other words, our gradient shift amounts to subtracting off the bulk equilibrium contribution when computing the average magnetic field. We conclude this brief survey by listing the boundary conditions we use in our Floquet calculations:
\begin{align}
    \left. \partial_x \theta'_1 \right|_{x=0,L} =& \, q \pm \sqrt{\frac{\beta}{1 + \beta}} \frac{d_1}{d_1 + d_2} \left. \left( \partial_t \theta'_1 + \partial_t \theta'_2 \right) \right|_{x=0,L} , \label{eq:radiative1}\\
    \left. \partial_x \theta'_2 \right|_{x=0,L} =& \, -\frac{q}{1 + \beta} \pm \sqrt{\frac{\beta}{1 + \beta}} \frac{d_2}{d_1 + d_2} \left. \left( \partial_t \theta'_1 + \partial_t \theta'_2 \right) \right|_{x=0,L} , \label{eq:radiative2}
\end{align}
where the $+$ ($-$) sign applies to the $x = 0$ ($x = L$) edge and we neglect the fast $E_p$ oscillations in Eq.~\eqref{eq:BCMulti1}. We note that, by taking $c \to c/n $ in Eq.~\eqref{eq:BCMulti1}, we can model propagation in an exterior medium with refractive index $n$, and the limit $n \to 0$ approaches Neumann boundary conditions. Throughout this work we use the value $n = 1$ unless stated otherwise.
\begin{figure*}[t!]
    \centering
    \includegraphics[scale = 1]{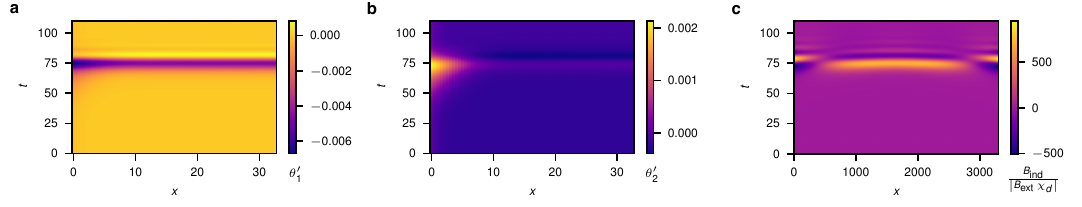}
    \caption{\textbf{Multi-bilayer spatio-temporal behaviors.} \textbf{a} Contour plot of the intra-bilayer phase $\theta'_1$, \textbf{b} contour plot of the inter-bilayer phase $\theta'_2$, and \textbf{c} contour plot of the average magnetic field induced within a unit cell, $B_{\rm{ind}}$, normalized by the equilibrium diamagnetic field amplitude, $|B_{\rm{ext}} \chi_d|$. Upon pumping, $\theta'_1$ and $\theta'_2$ monotonically grow in opposite directions (with $\theta'_1$ increasing negatively and $\theta'_2$ positively at the left edge), their profiles are mainly localized within a few penetration depths, but $\theta'_1$ develops an appreciable tail into the bulk of the segment. The magnetic field originates from edge currents, as explicitly addressed in the main text, and it propagates from the system's boundaries inwards at the speed of light. Oscillations at the edges of the segment represent the field emitted by the strip. These three plots are obtained for the parameter set (identical to that of Fig. 4 in the main text): $L=3300$, $V_0 = 3.5$, $q = 2.3 \times 10^{-4}$, $\alpha_1 = 0.4$, $\alpha_2 = 6$, and $\left| \chi_d \right| = 10^{-5}$. Panels (a) and (b) focus on the $0 \leq x \leq L/100 $ region.}
    \label{fig:SIFloquetContours}
\end{figure*}

\paragraph{Representative case.} In Fig.~\ref{fig:SIFloquetContours}(a), (b), and (c), we report contour plots illustrating the typical spatio-temporal behavior of, respectively, the intra-bilayer phase $\theta'_1$, the inter-bilayer phase $\theta'_2$, and the magnetic field ratio $B_{\rm{ind}} / |B_{\rm{ext}} \chi_d|$, see Eq.~\eqref{eq:ratioB}. The phases $\theta'_1$ and $\theta'_2$ are enhanced in opposite directions during the pump, with $\theta'_1$ growing negatively and $\theta'_2$ positively at the $x = 0$ boundary, which is the focus of Fig.~\ref{fig:SIFloquetContours}(a) and (b) (similar features, with opposite signs w.r.t. that at $x = 0$, dynamically emerge also at $x = L$). While their profiles are mostly concentrated within a few penetration depths, resembling the pre-solitonic excitations identified in the SG scenario, $\theta'_1$ develops a tail reaching further into the segment bulk. In Fig.~\ref{fig:SIFloquetContours}(c), we can appreciate how the magnetic field, which originates from the edge-instability, is established at the speed of light within the segment's interior. Our simulation also keeps track of the field emitted by the multi-bilayer system, as it is obtained through the radiative boundary conditions in Eqs.~\eqref{eq:radiative1} and~\eqref{eq:radiative2}, see the oscillations occurring at the two boundaries. Furthermore, we observe that, as a result of the field propagating symmetrically from both ends to the interior, a collision-like event unfolds around the center of the segment, leading to a standing-wave-type pattern. The magnitude of the latter profile is reasonably tied to the perfect geometry considered here, therefore we work with a phenomenological inter-bilayer damping coefficient $\alpha_2$ to account for leaks in the cavity to the radiation field formed by the region between bilayers. We consider a rather substantial value of $\alpha_2 = 6$ to suppress these oscillatory transients, with negligible influence on our peak magnetic response, see below.

\begin{figure*}[t!]
    \centering
    \includegraphics[scale = 1]{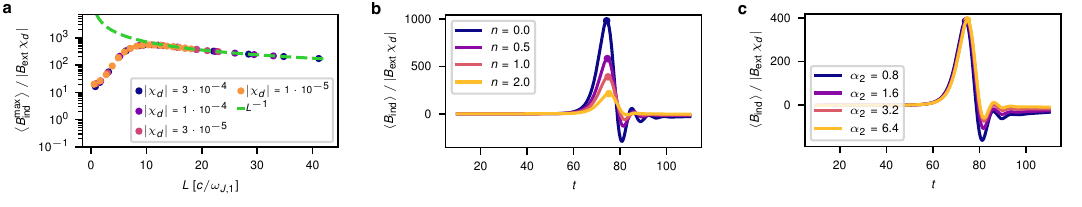}
    \caption{\textbf{Dependence on the $L$, $n$, and $\alpha_2$ parameters.} \textbf{a} Dependence of the peak magnetic response ratio to the equilibrium diamagnetic field, $\left\langle B^{\rm{max}}_{\rm{ind}} \right\rangle / |B_{\rm{ext}} \chi_d| $, where $\left\langle ... \right\rangle$ stands for segment averaging, versus the segment size $L$. The traveling wave character illustrated in Fig.~\ref{fig:SIFloquetContours}(c) suggests that, although the instability stems from the edges, it is not strictly proportional to the number of edges, i.e., inversely proportional to $L$. It instead shows a peculiar length dependence due to the precise response of the excited bulk modes. For very large segment sizes, the magnetic field pulse cannot travel all the way inside the bulk, and the effect is localized to the boundary in space, over a length-scale determined by either the finite pump duration or the dissipation of the bulk modes. Beyond this point, the generated flux saturates, and the average magnetic field decays as $ 1 / L $, see the green dashed line. The observed phenomenon is universal in $|\chi_d|$, as indicated by the accordance between the different sets of data points included the plot. Here the fixed parameters are: $V_0 = 3.5$, $\alpha_1 = 0.4$, and $\alpha_2 = 6$. \textbf{b} Time traces of the average induced magnetic field, normalized by the equilibrium diamagnetic field amplitude, for different values of the refraction index $n$ of the dielectric environment, with fixed parameters $L=3300 \, \lambda$, $V_0 = 3.5$, $q = 2.3 \times 10^{-4}$, $\alpha_1 = 0.4$, $\alpha_2 = 6$, and $\left| \chi_d \right| = 10^{-5}$. The dielectric environment is seen to influence the peak response, with the largest value obtained in the Neumann limit of $n \to 0$. \textbf{c} Time-traces of the average induced magnetic field for different values of the inter-bilayer damping parameter $\alpha_2$, with fixed parameters $L=3300 \, \lambda$, $V_0 = 3.5$, $q = 2.3 \times 10^{-4}$, $\alpha_1 = 0.4$, $ n = 1 $, and $\left| \chi_d \right| = 10^{-5}$. The inter-bilayer dissipation strength has little effect on the reported magnetic field build-up and maximum, but it suppresses the bounce-back oscillations we observe in the time profiles due to Fabry-P\'erot interferences.}
    \label{fig:SIFloquetDependence}
\end{figure*}

\paragraph{Effects of segment size, dielectric environment, and inter-bilayer damping.} Figure~\ref{fig:SIFloquetContours}(c) highlights the traveling-wave nature of the total magnetic field in the multi-bilayer strip. In Figure~\ref{fig:SIFloquetDependence}(a) we plot the length dependence of the ratio $\left\langle B^{\rm{max}}_{\rm{ind}} \right\rangle / |B_{\rm{ext}} \chi_d| $. Even though the effect is triggered at the edges of the sample, it is relatively system size independent due to the fact that the magnetic field enhancement travels to the bulk of the sample. Starting from a constant in small segments, the peak response increases with length. The propagation in the bulk is ultimately limited, either by the pump pulse duration or by dissipation of the photon modes. This gives a new characteristic length-scale localizing the propagating mode, $ l_{\rm{loc}} = c/t_{\rm{loc}} $ where $ t_{\rm{loc}} \sim 
\min \{\Delta t_{\rm{pulse}}, 1/\alpha_2\} $. Beyond this length-scale, the maximum induced flux saturates to constant value. As a result, the magnetic field averaged over the segment falls of as $1/L$, as confirmed by the green dashed line in our plot. Another important point highlighted by Fig.~\ref{fig:SIFloquetDependence}(a) is the universality in $|\chi_d|$. The accordance between the displayed data sets implies that $\left\langle B^{\rm{max}}_{\rm{ind}} \right\rangle / B_{\rm{ext}} $ is proportional to $|\chi_d|$, which is in turn proportional to the local superfluid density $n_s$. This is shown also Fig. 5(b) in the main text, and it reflects the fact that our dynamical response originates from the pump acting on the equilibrium diamagnetic screening properties, which are indeed encoded in the susceptibility $|\chi_d|$, see the Appendix~2.

We now turn to a description of Fig.~\ref{fig:SIFloquetDependence}(b), where we report time traces of $\left\langle B_{\rm{ind}} \right\rangle / |B_{\rm{ext}} \chi_d| $ for different values of the refraction index $n$, mimicking an interface to distinct dielectric environments. We observe that the overall magnitude of the system's response monotonically increases as the $n$ parameter goes down. Indeed, as compared to the result for our standard choice of $n = 1$ throughout this work (marked here in orange), the peak average magnetic field shows a factor of $2$ enhancement in the Neumann ($n \to 0$) limit, see the blue curve. Finally, in Fig.~\ref{fig:SIFloquetDependence}(c), we address the effect of dissipation between bilayers by displaying time traces of $\left\langle B_{\rm{ind}} \right\rangle / |B_{\rm{ext}} \chi_d| $ computed by varying the $\alpha_2$ coefficient. The latter quantity has minimal impact on the magnetic response's build-up and peak value. However, it plays a significant role in reducing the bounce-back oscillations observed as the system relaxes back to equilibrium.

\vspace{1.5cm}

\section{Appendix 6: Induced magnetic field profile beneath the mask}
\label{app6}

\begin{figure*}[t!]
    \centering
    \includegraphics[width = 0.6\textwidth]{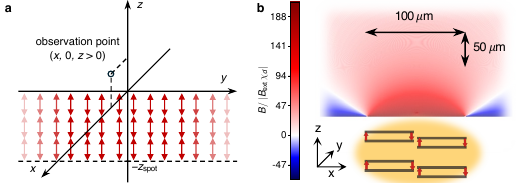}
    \caption{\textbf{Emitting edge currents and magnetostatic profile.} \textbf{a} Sketch of a current sheet, see the red arrows, extending over $-\infty < y < \infty$ and $-z_{\rm{spot}} < z < 0$. The sheet is located at $x = 0$. We compute the total magnetic field at the point $(x, y = 0, z > 0)$. \textbf{b} Magnetostatic field distribution in the region beneath the mask, computed for a two-segment system, demonstrating how the edge currents lead to a paramagnetic profile that can penetrate deep into the mask.}
    \label{fig:dipoles}
\end{figure*}

To estimate to what extent the magnetic field propagates from the pumped region to that beneath the mask, we make two main simplifying assumptions. First, the ensemble of illuminated edges is approximated as a 2D sheet of currents in the $y$-$z$ plane, localized at defects along the $x$-direction and vanishing for $z > 0$, where the mask blocks the pumping field, see Fig.~\ref{fig:dipoles}(a). Second, the medium beyond the mask's edge is effectively described, at THz frequencies, via a dielectric constant consistent with the experiment\cite{Sebastian24}.

Here, we consider the geometry shown in Fig.~\ref{fig:dipoles}(a). The mask is placed at $z \geq 0$, whereas for $-z_{\rm{spot}} < z < 0$ a single pumped edge at $x = 0$ is replaced by a sheet of dipole elements along the $z$-direction---extending over $-\infty < y < \infty$ by virtue of the above-assumed translational invariance across $y$---via a suitable averaging procedure over YBCO's unit cell. Our objective is to evaluate, at time $t$, the total magnetic field, along the $y$-direction, at an observation point $(x, y = 0, z > 0)$ representing the detector. The magnetic field contribution coming from a generic dipole in the sheet features both near-field and far-field components, decaying as $1 / R$ and $1 / R^2$, respectively, with $R$ being the distance from the dipole\cite{feynman1964lectures}. The far-field contribution is proportional to the time derivative of the electrical current $j_z$, whereas the near-field one is proportional to $j_z$ itself\cite{feynman1964lectures}. We note that, in our case, we expect a Gaussian-like behavior for the quantity $j_z$, implying the far-field contribution to be zero-averaging in time. We thus focus on the near-field result in the following. We mention, however, that similar formulas are readily derived for the far-field case as well. The $y$-component of the total magnetic field reads
\begin{equation}
    B(x, y = 0, z > 0, t) = \frac{\mu_0 x}{2 \pi} \int_{-\infty}^{0} d y_d \int_{-z_{\rm{spot}}}^{0} d z_d \frac{j_z (t - R / c)}{R^3} ,
    \label{eqn:Bdipoles}
\end{equation}
where $c$ is the speed of light in the material, $R = \sqrt{x^2 + y_d^2 + (z - z_d)^2}$ the distance between the observation point and the generic dipole element at location $(x = 0, y = y_d, z = z_d)$, and $j_z (t - R / c)$ the electrical current associated to each dipole, evaluated at the retarded time $t - R / c$ to account for the delayed contributions coming from the different sources in our geometry. To proceed further, we need to prescribe a time-dependence for the $j_z$ function. A first, interesting case we address is the impulsive limit: $c \, j_z(t - R/c) = j_0 \, \delta(t - R/c)$. Here, $j_0$ encodes the details of our YBCO averaging and is representative of the system's response in the pumped region, and $\delta$ stands for the Dirac delta function. By solving the double integral in Eq.~\eqref{eqn:Bdipoles}, we get
\begin{equation}
    B(x, 0, z > 0, t) =
\begin{cases} 
0, & \text{if } ct \leq \sqrt{x^2 + z^2} , \\
\frac{\mu_0 j_0 x}{2 \pi} \frac{1}{c^2 t^2} \left[ \arcsin \left( \frac{z + z_{\rm{spot}}}{\sqrt{c^2 t^2 - x^2}} \right) - \arcsin \left( \frac{z}{\sqrt{c^2 t^2 - x^2}} \right) \right], & \text{if } ct > \sqrt{x^2 + z^2} .
\end{cases}
\end{equation}
Rescaling the magnetic field by $\frac{\mu_0 j_0}{2 \pi a} \equiv B_{0} $ (a quantity representative of the illuminated YBCO overall response), position by $a = c \, t_0 = 100 \; \mu$m, and time by $t_0 = 1$ ps, we write
\begin{equation}
    \frac{B(x, 0, z > 0, t)}{B_{0}} =
\begin{cases} 
0, & \text{if } t \leq \sqrt{x^2 + z^2} , \\
\frac{x}{t^2} \left[ \arcsin \left( \frac{z + z_{\rm{spot}}}{\sqrt{t^2 - x^2}} \right) - \arcsin \left( \frac{z}{\sqrt{t^2 - x^2}} \right) \right], & \text{if } t > \sqrt{x^2 + z^2} .
\label{eqn:impulsive_norm}
\end{cases}
\end{equation}
The magnetic field profile for an arbitrary $j_z$ function of time can be obtained by convolution with the impulsive response in Eq.~\eqref{eqn:impulsive_norm}. In the main text, we work with a multiple-sheet system, showing the results for a Gaussian-like profile centered around $t = 0$, with a standard deviation of $0.5$, choosing $x = 0.4$, $z_{\rm{spot}} = 1$, and various $z \sim 1$ values. More specifically, for the two-segment example we consider a profile of the form $j_z = \sum_{\lambda} j_0 \, \delta(x - x_{\lambda}) \, (-1)^{\lambda+1} \, \theta(-z) \, \theta(z + z_{\rm{spot}}) \, \exp{[ -t^2 / ( 2 \sigma_t^2 ) ]}$, where even (odd) $\lambda$ annotates right (left) edges, with $\sigma_t = 0.5$, $x_1 = 0, x_2 = L_{\rm{seg}}, x_3 = L_{\rm{seg}} + \delta x $, $x_4 = 2 L_{\rm{seg}} + \delta x$, $L_{\rm{seg}} = 0.5$ (segments' length), and $\delta x = 0.01$ (separation between the two segments). The framework we employed compares reasonably well with the experimental counterpart\cite{Sebastian24}.

We conclude by addressing the result obtained upon convolving the impulsive response with a constant function---what we call the magnetostatic limit. In normalized units, we obtain the exact expression
\begin{equation}
    \frac{B(x, 0, z > 0)}{B_{0}} = \arctan \left( \frac{z + z_{\rm{spot}}}{x} \right) - \arctan \left( \frac{z}{x} \right) .
\end{equation}
A similar formula is readily obtained for single and multiple YBCO segments (which involve a number of dipole sheets) by superimposing single-sheet magnetic field contributions. In Fig.~\ref{fig:dipoles}(b), we show the magnetostatic result for a two-segment system.

\bibliography{references}{}

\end{document}